\documentclass[preprint,preprintnumbers,amsmath,amssymb,superscriptaddress,nofootinbib]{revtex4}
\usepackage{graphicx,color}
\usepackage{amsmath,amssymb}
\usepackage{url}
\usepackage{epstopdf}

\newcommand{\hs}{\hspace*{0.5cm}}

\newcommand{\be}{\begin{equation}}
\newcommand{\ee}{\end{equation}}
\newcommand{\bea}{\begin{eqnarray}}
\newcommand{\eea}{\end{eqnarray}}
\newcommand{\nn}{\nonumber}
\newcommand{\crn}{\nonumber \\}

\newcommand{\al}{\alpha}

\newcommand{\fr}{\frac}
\newcommand{\bc}{\begin{center}}
\newcommand{\ec}{\end{center}}

\newcommand {\ba}{\begin{array}}
\newcommand {\ea}{\end{array}}
\newcommand{\ben}{\begin{enumerate}}
\newcommand{\een}{\end{enumerate}}

\usepackage{epsfig,graphicx}
\usepackage{bm}
\usepackage{dcolumn}

\begin{document}

\preprint{}

\title{Lepton flavor violating decays of the SM-like Higgs boson $h\rightarrow e_ie_j$, and $e_i \rightarrow e_j\, \gamma $ in a flipped 3-3-1 model}

\author{T.~T.~Hong}
\affiliation{An Giang University, VNU - HCM, 18 Ung Van Khiem Street, Long Xuyen,  An Giang 880000, Vietnam}

\author{H.~T.~Hung}
\affiliation{Department of Physics, Hanoi Pedagogical University 2, Phuc Yen,  Vinh Phuc 280000, Vietnam}

\author{H.~H.~Phuong}
\affiliation{Department of Physics, Hanoi Pedagogical University 2, Phuc Yen,  Vinh Phuc 280000, Vietnam}
%
%
\author{L.~T.~T.~Phuong}
\affiliation{An Giang University, VNU - HCM, 18 Ung Van Khiem Street, Long Xuyen,  An Giang 880000, Vietnam}

\author{L.T. Hue }\email{lethohue@duytan.edu.vn (Corresponding~author)}
\affiliation{ Institute for Research and Development, Duy Tan University,  Da Nang 550000, Vietnam
}

\begin{abstract} 
In the framework of the flipped 3-3-1 model introduced recently~\cite{Fonseca:2016tbn},   the lepton-flavor-violating (LFV) decay $\mu \rightarrow  3e$  was predicted to  have a large branching ratio (Br) close to the recent experimental limit.  We will show  that the   Br of LFV decays of the standard-model-like (SM-like) Higgs boson decays (LFVHD) Br$(h\rightarrow e_ae_b)$ may also be large. Namely, the Br$(h\rightarrow \mu\tau,e\tau)$ can reach values of   $\mathcal{O}(10^{-4})-\mathcal{O}(10^{-5})$, which will be reach the upcoming experimental sensitivities.  On the other hand,  for LFV decays of charged leptons (cLFV)  $(e_b\rightarrow e_a\gamma)$, the branching ratios  are well below  experimental bounds.   
\end{abstract}

\pacs{
 }
\maketitle
\section{Introduction}
\allowdisplaybreaks

Since the SM-like Higgs boson was discovered in 2012~\cite{Aad:2012tfa,Chatrchyan:2012xdj,Chatrchyan:2013lba}, the LFV decays of this Higgs boson have been sought  for by experiments at the Large Hadron Collider (LHC)~\cite{Khachatryan:2015kon, Aad:2015gha, Khachatryan:2016rke, Aad:2016blu, Sirunyan:2017xzt, Aad:2019ugc}.  Recent experimental lower bounds on the LFV decays of the standard-model-like (SM-like) Higgs decays (LFVHD) $h\rightarrow e_be_a$ are
\begin{align}
\label{eq_hebax}
\mathrm{Br}(h\rightarrow \tau\mu), \mathrm{Br}(h\rightarrow \tau e)&\leq  \mathcal{O}(10^{-3}),\crn
\mathrm{Br}(h\rightarrow \mu e)&<  3.5\times 10^{-4}. 
\end{align}
An updated lower bound  $ \mathrm{Br}(h\rightarrow \mu e)<6.1\times 10^{-5}$ has been reported recently by the ATLAS  Collaboration~\cite{Aad:2019ojw}. Recent studies have predicted that  lower bounds from experiments  for Br$(h\rightarrow\mu\tau,e\tau)$ could reach the orders  of $\mathcal{O}(10^{-4})-\mathcal{O}(10^{-5})$~\cite{Banerjee:2016foh, Chakraborty:2016gff, Chakraborty:2017tyb, Qin:2017aju}. 

The LFVHD has been studied in many  models beyond the SM, from seesaw and inverse seesaw models~\cite{Pilaftsis:1992st, Korner:1992zk, Arganda:2014dta, Thao:2017qtn, Marcano:2019rmk} to  more complicated ones~\cite{ DiazCruz:1999xe, Harnik:2012pb, Falkowski:2013jya, Celis:2013xja, Dery:2014kxa, He:2015rqa,Dorsner:2015mja, Heeck:2014qea, Crivellin:2015lwa, deLima:2015pqa, Omura:2015nja, Campos:2014zaa, Crivellin:2015mga, Das:2015zwa, Lami:2016mjf, Omura:2015xcg, Altmannshofer:2015esa, Chang:2016ave, Chen:2016lsr, Huitu:2016pwk, Cheung:2015yga, Bizot:2015qqo, Aoki:2016wyl, Guo:2016ixx, Hue:2015fbb, Thuc:2016qva, Phan:2016ouz, Herrero-Garcia:2017xdu, Yang:2016hrh, Cai:2017jrq, Arganda:2017vdb, Nguyen:2018rlb, Chamorro-Solano:2017toq,  Vicente:2019ykr}, including the suppersymmetric versions~\cite{Brignole:2003iv, DiazCruz:2002er, Brignole:2004ah, Arganda:2004bz, Giang:2012vs, Arana-Catania:2013xma, Binh:2013axa, Arganda:2015naa, Arganda:2015uca, Baek:2015fma, Baek:2015mea, Zhang:2015csm, Chattopadhyay:2019ycs}.  Many of these models predict very lagre Br$(h\rightarrow \tau\mu,\tau e)$ with the order of $\mathcal{O}(10^{-5})$, implying that  LFVHD decays will be signals as new physics that will be tested experimentally in the near future.

The models beyond the SM constructed by extending the gauge group $SU(3)_C\times SU(2)_L\times U(1)_Y$ into the group $SU(3)_C\times SU(3)_L\times U(1)_X$ (3-3-1) models  may predict large LFV decay branching ratios. This  can be explained based on the common property of the popular 3-3-1 models~\cite{Singer:1980sw, Pisano:1991ee, Frampton:1992wt, Foot:1994ym, Montero:1992jk}  that   left-handed fermions are usually arranged into $SU(3)_L$ (anti)triplets. Hence,  there will appear  couplings of new heavy leptons in the third components of these lepton representations with normal charged leptons and gauge or Higgs bosons. The mixing of these heavy leptons is an important source of LFV mediation at the one-loop level. Therefore, LFV decays of charged leptons in the framework of  3-3-1 models have been widely investigated~\cite{Liu:1993gy, Boucenna:2015zwa, Arcadi:2017xbo, Lindner:2016bgg, Hue:2017lak, Hue:2013uw, Long:2018dun, CarcamoHernandez:2019vih}.  Many of the 3-3-1 models can explain the recent lower bounds on the decays Br$(e_b\rightarrow e_a\gamma)$~\cite{TheMEG:2016wtm, Aubert:2009ag}
\begin{align}
\label{eq_ebagaex}
\mathrm{Br}(\tau\rightarrow \mu\gamma)&<4.4\times 10^{-8},\crn
\mathrm{Br}(\tau\rightarrow e\gamma)&<3.3\times 10^{-8},\crn
\mathrm{Br}(\mu\rightarrow e\gamma)&< 4.2\times 10^{-13}.
\end{align}
In future projects, new sensitivities  for these decay channels  will be $\mathrm{Br}(\mu\rightarrow e\gamma) \sim \mathcal{O}(10^{-14})$~\cite{Baldini:2013ke} and $\mathrm{Br}(\tau\rightarrow \mu\gamma,e\gamma) \sim \mathcal{O}(10^{-9})$~\cite{Aushev:2010bq}. They will be used to determine  allowed regions of the parameter spaces of the 3-3-1 models  for further  studying other LFV decays such as those of the SM-like Higgs boson $h\rightarrow e_b^{\pm}e_a^{\mp}$.  They just have been investigated in just a few specific 3-3-1 models~\cite{Hue:2015fbb, Nguyen:2018rlb}, where the LFV sources come from the mixing of heavy neutrinos. In particular, the 3-3-1 model with inverse seesaw neutrinos~\cite{Nguyen:2018rlb} predicts  very small regions of parameter space that give large Br$(h\rightarrow \tau\mu,\tau e)\simeq O(10^{-5})$ and also  satisfy  the current bounds of Br$(\mu\rightarrow e\gamma)$.  Recently,  an interesting   flipped 3-3-1model  has been constructed~\cite{Fonseca:2016tbn}, where the left-handed lepton is arranged in a lepton sextet, while the left-handed $\tau$ and $\mu$ are still the same as those known previously. In addition,   all left-handed quarks are also arranged in the same $SU(3)_L$ triplets so that the model is anomaly free. The treel-level flavor neutral changing currents caused by the heavy neutral boson $Z'$ do not appear; hence $m_{Z'}$ is not constrained by the corresponding experimental data. The active neutrino and electron masses can be produced consistent with experiments through loop corrections~\cite{Fonseca:2016xsy}. The effect of the Higgs sextet on fermion and Higgs boson couplings was discussed in ref.~\cite{Dias:2006ns}. The Higgs potentials relating to the Higgs sextets were  studied in refs.~\cite{Diaz:2004fs, Diaz:2003dk}.  Based on these ingredients, our aim in this work is to  investigate the LFV decays of charged leptons $e_b\rightarrow e_a \gamma$ and the SM-like Higgs  boson $h\rightarrow e_be_a$  in the framework of the  flipped 3-3-1 model. 

 Our work is arranged as follows. In  Sects~\ref{models} and \ref{sec_boson}, we will collect the main content of the flipped 3-3-1 model, where masses, physical states, and needed couplings for calculating branching ratios of the LFV decays are presented.  The analytic formulas of LFV branching ratios and the corresponding numerical investigations will be shown in section \ref{sec_LFVdecay}. We will summary main results in Sect~\ref{sec_conclusion}. Finally, there are two appendices showing the details of the one loop formulas contributing to the LFV decays of charged lepton (cLFV) amplitudes of the decays $e_b \rightarrow e_a\gamma$ and the  equations for minimal conditions of the Higgs potential considered in this work.

\section{The flipped 3-3-1 model}
\label{models}
\subsection{The model review }
We follow the model introduced in Ref.~\cite{Fonseca:2016tbn}, where the particle content  is presented in Table~\ref{tab:Fields_of_new_model}. 
\begin{table}[ht]
	\centering
		\begin{tabular}{ccccc}
			\toprule 
			Name & 331 rep. & SM group decomposition & Components & \# flavors
			\tabularnewline
			$L_{e}$ & $\left(\mathbf{1},\mathbf{6},-\frac{1}{3}\right)$ & $\left(\mathbf{1},\widehat{\mathbf{3}},0\right)+\left(\mathbf{1},\widehat{\mathbf{2}},-\frac{1}{2}\right)+\left(\mathbf{1},\widehat{\mathbf{1}},-1\right)$ & $\left(\begin{array}{ccc}
			\left( \Sigma^{-}_R\right)^c & \frac{1}{\sqrt{2}}\Sigma^{0}_L & \frac{1}{\sqrt{2}}\nu_{eL}\\
			\frac{1}{\sqrt{2}}\Sigma^{0}_L & \Sigma^{-}_L & \frac{1}{\sqrt{2}}e_L\\
			\frac{1}{\sqrt{2}}\nu_{eL} & \frac{1}{\sqrt{2}}e_L & E_{eL}
			\end{array}\right)$ & 1
			\tabularnewline
			$L_{\alpha=\mu,\tau}$ & $\left(\mathbf{1},\mathbf{3},-\frac{2}{3}\right)$ & $\left(\mathbf{1},\widehat{\mathbf{2}},-\frac{1}{2}\right)+\left(\mathbf{1},\widehat{\mathbf{1}},-1\right)$ & $\left(\nu_{\alpha},e_{\alpha},E_{\alpha}\right)^{T}_L$ & 2\tabularnewline
			$e_{\alpha {R}}$ & $\left(\mathbf{1},\mathbf{1},-1\right)$ & $\left(\mathbf{1},\widehat{\mathbf{1}},-1\right)$ & $e_{\alpha{R}}$ & 6\tabularnewline
			$Q_{\alpha}$ & $\left(\mathbf{3},\overline{\mathbf{3}},\frac{1}{3}\right)$ & $\left(\mathbf{3},\widehat{\mathbf{2}},\frac{1}{6}\right)+\left(\mathbf{3},\widehat{\mathbf{1}},\frac{2}{3}\right)$ & $\left(d_{\alpha},-u_{\alpha},U_{\alpha}\right)^{T}_{L}$ & 3\tabularnewline
			$u_{\alpha{R}}$ & $\left(\mathbf{3},\mathbf{1}, \frac{2}{3}\right)$ & $\left(\mathbf{3},\widehat{\mathbf{1}}, \frac{2}{3}\right)$ & $u_{\alpha {R}}$ & 6\tabularnewline
			$d_{\alpha{R}}$ & $\left( \mathbf{3},\mathbf{1}, -\frac{1}{3}\right)$ & $\left( \mathbf{3},\widehat{\mathbf{1}}, -\frac{1}{3}\right)$ & $d_{\alpha{R}}$ & 3\tabularnewline
			$\phi_{i=1,2}$ & $\left(\mathbf{1},\mathbf{3},\frac{1}{3}\right)$ & $\left(\mathbf{1},\widehat{\mathbf{2}},\frac{1}{2}\right)+\left(\mathbf{1},\widehat{\mathbf{1}},0\right)$ & $\left(H_{i}^{+},H_{i}^{0},\sigma_{i}^{0}\right)^{T}$ & 2\tabularnewline
			$\phi_{3}$ & $\left(\mathbf{1},\mathbf{3},-\frac{2}{3}\right)$ & $\left(\mathbf{1},\widehat{\mathbf{2}},-\frac{1}{2}\right)+\left(\mathbf{1},\widehat{\mathbf{1}},-1\right)$ & $\left(H_{3}^{0},H_{3}^{-},\sigma_{3}^{-}\right)^{T}$ & 1\tabularnewline
			$S$ & $\left(\mathbf{1},\mathbf{6},\frac{2}{3}\right)$ & $\left(\mathbf{1},\widehat{\mathbf{3}},1\right)+\left(\mathbf{1},\widehat{\mathbf{2}},\frac{1}{2}\right)+\left(\mathbf{1},\widehat{\mathbf{1}},0\right)$ & $\left(\begin{array}{ccc}
			\Delta^{++} & \frac{1}{\sqrt{2}}\Delta^{+} & \frac{1}{\sqrt{2}}H_{S}^{+}\\
			\frac{1}{\sqrt{2}}\Delta^{+} & \Delta^{0} & \frac{1}{\sqrt{2}}H_{S}^{0}\\
			\frac{1}{\sqrt{2}}H_{S}^{+} & \frac{1}{\sqrt{2}}H_{S}^{0} & \sigma_{S}^{0}
			\end{array}\right)$ & 1\tabularnewline
		\end{tabular}
		\caption{\label{tab:Fields_of_new_model}Representations for the flipped 3-3-1 model, taken from~Ref.~\cite{Fonseca:2016tbn}, the notations of fermions are  Dirac spinors.}
\end{table}
All fermions  are written in terms of Dirac spinors. 

The electric charge  operator is:
\begin{equation}\label{eq_electricQ}
Q=T^3 + \frac{1}{\sqrt{3}} T^8 +X,
\end{equation} 
where $T^{3,8}$ are diagonal generators of the $SU(3)$ group.

 These Higgs bosons  develop vacuum 
expectation values (VEV) defined  as 
\bea &&\sigma^0_i= n_i +\frac{1}{\sqrt{2}} \left(R_{\sigma_i}+i I_{\sigma_i}\right), \hs \langle \sigma^0_i\rangle=n_i ,\; i=1,2, S,
\crn &&  H^0_\alpha = k_\alpha + \frac{1}{\sqrt{2}} \left(R_{\alpha}+ iI_{\alpha}\right), \hs \langle H_{\alpha}^0 \rangle= k_{\alpha} ,\; \alpha=1,2,3, S, \crn
&&\Delta^0=\epsilon_S + \frac{1}{\sqrt{2}} \left(+R_{\Delta}+ iI_{\Delta}\right), \hs \langle \Delta^0\rangle= \epsilon_S,
 \label{neuHiggs}\eea
where  $\epsilon_S\ll\, k_{1,2,3,S}\ll \,n_{1,2,S}$ in  general~\cite{Fonseca:2016tbn}. In addition, it was shown that $\epsilon_S$ and $k_S$ should be small to successfully generate  neutrino mass consistent with experimental data.   Hence, we can take $k_s=\epsilon_S\simeq0$ when solving the masses and physical states of Higgs and gauge bosons. 

The Yukawa Lagrangian for the lepton sector is 
\begin{equation}\label{eq_YukawaLepton}
-\mathcal{L}^Y_{\mathrm{lepton}}=\sum_{i=1}^2 \sum_{\alpha=\mu,\tau}\sum_{\beta=1}^6y^{\ell(i)}_{\alpha\beta}\overline{e_{\beta{R}}} L_{\alpha}\phi^*_i +\sum_{\beta=1}^6y^{\ell\prime}_{\beta} \overline{e_{\beta{R}}}L_e S^* +y^{\ell\prime\prime} \overline{(L_e)^c}L_eS +\mathrm{h.c.},
\end{equation}
where the  invariant term  of the tensor product of the three sextets is expanded as  $\overline{(L_e)^c}L_eS=\epsilon_{abc}\epsilon_{ijk}\overline{(L_e)^c}_{ai}(L_e)_{bj}S_{ck}$~\cite{Liu:1993gy,DeConto:2015eia},  $(L_e)^c_{ai}\equiv C\overline{(L_e)_{ai}}^T$.  Note that $\phi_3$  only appears in the Yukawa part of the quark.

The fermions are presented as two-component spinors in the original version; see table~\ref{tab:Fields_of_new_model} in Ref.~\cite{Fonseca:2016tbn}. In this work, we will use the Dirac (four-component) spinor notation,  based on the equivalence given in detail in Ref.~\cite{Dreiner:2008tw}. In particular, a Dirac spinor $f=(f_L,\; f_R)^T$, where $f_{L,R}$ is  the respective left (right) component of a Dirac fermion, namely $f_L=P_Lf$ and $f_R=P_Rf$. The Dirac conjugation is $\overline{f}=f^{\dagger}\gamma^0=(\overline{f_R},\; \overline{f_L})$. The charge conjugation is $f^C\equiv C\overline{f}^T=((f_R)^c, (f_L)^c)^T$, implying that  $(f_{R,L})^c=P_{L,R}f^C$. A Majorana fermion satisfying  $f^C=f$ results in $f_{L,R}=(f_{R,L})^c$. The mass term of all fermions at tree level is
\begin{align}
\label{eq_LeptonMass}
-\mathcal{L}^{\mathrm{lepton}}_{\mathrm{mass}}&= \sum_{i=1}^2 \sum_{\alpha=\mu,\tau}\sum_{\beta=1}^6y^{\ell(i)}_{\alpha\beta}\overline{e_{\beta{R}}}\left( e_{\alpha L} k_i + E_{\alpha L} n_i\right) + \sum_{\beta=1}^6y^{\ell\prime}_{\beta} \overline{e_{\beta{R}}} \left( \Sigma^-_L\epsilon_S  + e_L k_S  +E_{eL} n_S \right)  \crn 
&+y^{\ell\prime\prime}  \left[ 2\epsilon_S \left(\overline{\Sigma^-_R} E_{eL} - \overline{(\nu_{eL})^c} \nu_{eL}\right) + 2 k_S \left(-\overline{\Sigma^-_R} e_L + \frac{1}{\sqrt{2}} \overline{(\nu_{eL})^c} \Sigma^0_{L}\right) \right.  \crn
&\left.+ n_S \left( 2\overline{\Sigma^-_R} \Sigma^-_L - \overline{( \Sigma^0_{L})^c} \Sigma^0_{L}\right) \right]+\mathrm{h.c.},
\end{align}
where we have used the identity $\overline{\psi^c_a} \psi^c_b=\overline{\psi_b} \psi_a$ for leptons. 

\begin{figure}[ht]
	\centering
	\begin{tabular}{cc}
		\includegraphics[width=5cm]{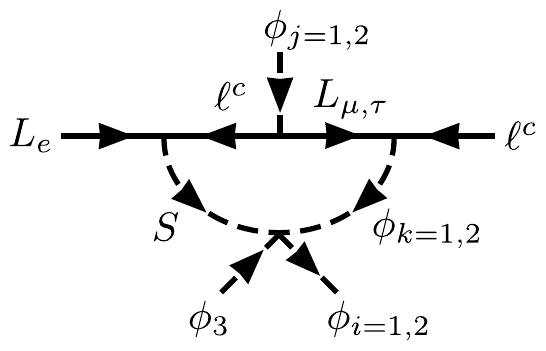}& \includegraphics[width=5cm]{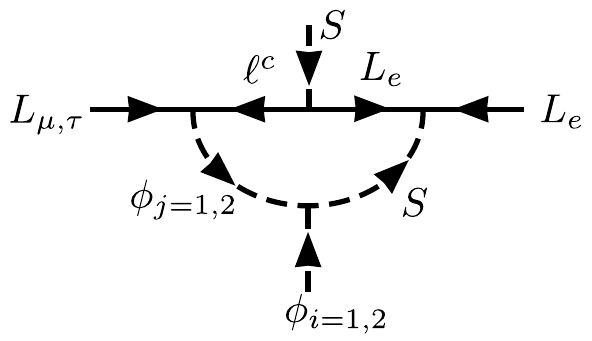}
		\\
	\end{tabular}
	\caption{Feynman diagrams giving one-loop corrections to the masses of electron (left panel) and active neutrinos (right panel)~\cite{Fonseca:2016tbn}, where $\ell^c$ means $\ell^c_{\alpha}\equiv (e_{\alpha R})^c$}\label{fig_MassCorrection}
\end{figure}

According to the discussion on Ref.~\cite{Fonseca:2016tbn}, in the basis $\Psi^{\ell}_{L,R}=(e_{\alpha},\; E_{\alpha},\; E_{e},e,\Sigma^-)^T_{{L,R}}$ the mass matrix of charged leptons always has one massless eigenstate at tree level, corresponding to the normal electron mass $m_e=0$. This is also the case  for active neutrinos. However,  when the loop corrections are included, the consistent masses of electrons and active neutrinos are obtained. The one-loop Feynman diagrams  corresponding to these corrections are  given in Fig.~\ref{fig_MassCorrection}, and were pointed out in Ref.~\cite{Fonseca:2016tbn},  
along with a very detailed discussion on this property of the flipped 3-3-1 model.  Accordingly, using the minimal Higgs sector given in Table~\ref{tab:Fields_of_new_model}, the experimental data  of an inverse hierarchy for active neutrinos can be fitted. Adding more scalar fields to the model will be another way to solve the problem of the neutrino oscillations that can be fitted with recent experimental data. As we will show, this problem does not affect significantly our discussion on LFV decays.

Because loop corrections are needed to generate masses of only very light leptons, namely electrons and active neutrinos, the other corrections to the lepton mass matrices are also  smaller than other heavy masses  appearing in the model. This is also because of  another reason that one-loop  corrections are  suppressed by the two factors $1/(16\pi^2)$ and $1/M^2$ relating respectively to the  one-loop integral and new heavy masses $M$ of a new particle running in the loop. In conclusion, loop corrections  make tiny contributions to the lepton mass matrices. Hence,  we will ignore loop corrections to the masses of heavy particles from now on.

 For simplicity, in this work  we will  assume  that only exotic charged leptons $E_e, E_{\mu}, E_{\tau}$ mix with each other to guarantee the existence of LFV couplings that contribute to one-loop amplitudes of the LFV decays. On the other hand,  all of the original states of the SM charged leptons and $\Sigma^-$ are physical. This corresponds to the condition that $\epsilon,k_S,n_1,k_2\simeq0$. The large Yukawa couplings  of the physical states $\mu,\tau$ and $\Sigma^-$ are 
\begin{align}
\label{eq_nonzeroYlepton}
y^{\ell(1)}_{11}&= \frac{m_{\mu}}{k_1}, \quad y^{\ell(1)}_{22}= \frac{m_{\tau}}{k_1},  \quad\,  y^{\ell \prime\prime}= \frac{m_{\Sigma^-}}{2n_S}. 
\end{align}
Note that the  masses of electrons and active neutrinos come  from loop corrections.

The original basis $(E_{\mu}, E_{\tau},E_{e})$ corresponds  to the following mass term: 
 \begin{align}
 \label{eq_massEmutaue}
 -\mathcal{L}^{E}_{mass}&=\begin{pmatrix}
 \overline{E_{\mu R}}&  \overline{E_{\tau R}} &   \overline{E_{eR}}\end{pmatrix}  \mathcal{M}_{E_{\mu,\tau,e}} \begin{pmatrix} E_{\tau L}&   E_{\mu L} &  E_{e L}
 \end{pmatrix}^T +h.c., \crn 
 \mathcal{M}_{E_{\mu,\tau,e}}&=n_2\begin{pmatrix}
 y^{\ell (2)}_{13}&  y^{\ell (2)}_{14}& y^{\ell (2)}_{15} \\ 
 y^{\ell (2)}_{23}& y^{\ell (2)}_{24} & y^{\ell (2)}_{25} \\ 
\frac{n_S}{n_2}y^{\ell\prime}_3 & \frac{n_S}{n_2}y^{\ell\prime}_4 & \frac{n_S}{n_2}y^{\ell\prime}_5. 
 \end{pmatrix} \equiv \,n_2Y^{\ell},
 \end{align}
 where we have used the assumption that some of the Yukawa couplings in the Lagrangian~\eqref{eq_YukawaLepton} are zeros.
 The lepton mass matrix in Eq.~\eqref{eq_massEmutaue} is arbitrary;  hence it is diagonalized by the following transformation: 
 \begin{align}
 \label{eq_E123}
 V^{E\dagger}_R \mathcal{M}_{E_{\mu,\tau,e}}V^E_L&=\mathrm{diag}(m_{E_1},m_{E_2},m_{E_3}), \crn 
\begin{pmatrix}
E_{\mu} \\ 
E_{\tau}
\\ 
E_{e} 
\end{pmatrix}_{R,L} &=V^E_{R,L}\begin{pmatrix}
E_{1} \\ 
E_{2}
 \\ 
E_{3} 
 \end{pmatrix}_{R,L},
 \end{align}
 where $m_{E_{i}}$ are masses of the physical states $E_{iL(R)}$, $i=1,2,3$.   For simplicity,  in this work  we will choose $V^E_R=I_3$, while $V^E_L$ is parameterized  in terms of three free mixing angles $\theta^E_{ij}$, $i,j=1,2,3$ ($i<j$), namely 
 \bea V^E_{L}&\equiv &\begin{pmatrix}
 	1	& 0 &0  \\ 
 	0	&c^E_{23}  &s^E_{23}  \\ 
 	0&  	-s^E_{23}& c^E_{23}
 \end{pmatrix}\,\begin{pmatrix}
 	c^E_{13}	& 0 &s^E_{13}  \\ 
 	0	&1  &0  \\ 
 	-s^E_{13}&  0& c^E_{13}
 \end{pmatrix}\,\begin{pmatrix}
 	c^E_{12}	& s^E_{12} &0  \\ 
 	-s^E_{12}	&c^E_{12}  &0  \\ 
 	0& 0 	&1
 \end{pmatrix} \crn
 &=&\left(
 \begin{array}{ccc}
 	c^E_{12} c^E_{13} & c^E_{13} s_{12} &  s^E_{13}\\
 	-c^E_{23} s^E_{12} - c^E_{12} s^E_{13} s^E_{23} & c^E_{12} c^E_{23}- s^E_{12} s^E_{13} s^E_{23} & c^E_{13} s^E_{23} \\
 	s^E_{12} s^E_{23}- c^E_{12} c^E_{23} s^E_{13} & -c^E_{12} s^E_{23} -  c^E_{23}  s^E_{12} s^E_{13} & c^E_{13} c_{23} \\
 \end{array}
 \right), \label{eq_VEL}\eea 
 where $s_{ij}\equiv\sin\theta^E_{ij}$, $c^E_{ij}\equiv\cos\theta^E_{ij}$, and all Dirac and Majorana phases are set to be zeros.  This matrix exactly satisfies  the unitary property. We will use  $s^E_{ij}$ as free parameters.  

Other Yukawa couplings are non-zero for generating active neutrino masses and mixing consistent with experiments (see discussions in ref.~\cite{Fonseca:2016tbn}), but they are assumed to be suppressed in this work.  We also note that the conditions in Eq.~\eqref{eq_nonzeroYlepton}  still allow right SM quark masses and mixing consistent with experimental data.  Similarly, there is one heavy Majorana neutrino $\Sigma_M=(\Sigma^0,\; \Sigma^{0\dagger})^T$ with the mass term  $-1/2(-2y^{\ell\prime\prime}n_S)\Sigma^0\Sigma^0 +\mathrm{h.c.}$. Three other active neutrinos get consistent masses and mixing from loop corrections, which prefer the inverted order of active neutrino data oscillation. Their physical states are denoted as $n_1,n_2,n_3$~\cite{Fonseca:2016tbn}. The masses and  mass eigenstates of heavy neutral leptons  are 
\begin{equation}\label{eq_Sima0mass}
n_4=i\Sigma_M,\quad m_{n_4}=m_{\Sigma^-}=2n_S y^{\ell\prime\prime}. 
\end{equation}
Yukawa coupling terms in the Lagrangian~\eqref{eq_YukawaLepton} containing normal charged leptons  are  
\begin{align}
\label{eq_SMchargedlep} 
\mathcal{L}^{\mathrm{Y}}_{\ell}=& -\frac{m_{\mu}}{k_1} \left[ H^{0*}_1\overline{\mu_R} \mu_L +  \sigma^{0*}_1\overline{\mu_R} E_{\mu L}  + H^{-}_1\overline{\mu_R} \nu_{\mu L} \right] \crn
& - \frac{H^{0*}_2}{\sqrt{2}} \left[ \overline{E_{\mu R}}y^{\ell(2)}_{13} +  \overline{E_{\tau R}}y^{\ell(2)}_{14} + \overline{E_{e R}}y^{\ell(2)}_{15} \right]\mu_L   \crn 
&-  \frac{m_{\tau}}{k_1} \left[ H^{0*}_1\overline{\tau_R} \tau_L +  \sigma^{0*}_1\overline{\tau_R} E_{\tau,L}  + H^{-}_1\overline{\tau_R} \nu_{\tau,L} \right] \crn
& - \frac{H^{0*}_2}{\sqrt{2}}\left[ \overline{E_{\mu,R}}y^{\ell(2)}_{23} +  \overline{E_{\tau,R}}y^{\ell(2)}_{24} + \overline{E_{e,R}}y^{\ell(2)}_{25} \right]\tau_L   \crn 
&- \frac{H^{0*}_S}{\sqrt{2}}  \left[ \overline{E_{\mu R}}y^{\ell\prime}_3 + \overline{E_{\tau R}}y^{\ell\prime}_4+ \overline{E_{eR}}y^{\ell\prime}_5\right]\,e_L +\frac{m_{\Sigma^-}}{n_S} H^{0*}_S \overline{\Sigma^-_{R}}\,e_L \crn 
&-\frac{m_{\Sigma^-}}{\sqrt{2} n_S} \Delta^+ \nu_{e,L}\,e_L  + \frac{m_{\Sigma^-}}{n_S} \Delta^{++} \overline{(e_L)^c}e_L+ \frac{im_{\Sigma^-}}{\sqrt{2}n_S} H^+_S n_{4,L}\,e_L+\mathrm{h.c.}
\end{align}
Corresponding to the above assumption that all charged leptons are diagonal, Yukawa couplings  relating to one-loop corrections must guarantee that new Higgs bosons should couple to different SM charged leptons. As we will show later, the SM-like Higgs bosons will be $h\simeq R_3$ when we assume that $k_1\ll k_3$.  Combined with Lagrangian~\eqref{eq_SMchargedlep}, we can see that tree-level couplings of the SM-like Higgs boson $h \overline{e_i}e_j$ do not appear. The heavy neutral lepton $n_4$ does not couple with normal charged leptons. The couplings  $h \overline{e_i}e_i$ appear from the small mixing of $R_3$ and $R_1$ for $e_i=\mu,\tau$ and loop corrections for the electron.  These couplings have small effects on the LFV decays so we omit them from now on.

After breaking, the masses and physical states of all gauge bosons are determined as follows.

 \section{ \label{sec_boson} Higgs and gauge bosons} 
\subsection{Gauge boson}
The covariant derivative  of the $SU(3)_L\times U(1)_X$ is defined as
\begin{equation}\label{eq_Dsu3}
D_{\mu}\equiv \partial_{\mu} -igW^a_{\mu}T^a -ig_X T^9XX_{\mu},
\end{equation}
where $T^a$ ($a=1,2,..,8$) is the $SU(3)$ generator with respective gauge boson $W^a_{\mu}$, $T^9=\frac{I}{\sqrt{6}}$ is the $U(1)_X$ generator with the gauge boson $X_{\mu}$, and $X$ is the $U(1)_X$ charge of the field acted by the covariant derivative.  The  particular forms of the generators are:
\begin{itemize}
	\item For an $SU(3)_L$ singlet: $T^a=0~\forall a=1,2,..,8$, $T^9=\frac{1}{\sqrt{6}}$.
	\item For an $SU(3)_L$ triplet: $T^a=\frac{1}{2}\lambda_a~\forall a=1,2,..,8$, $T^9=\frac{1}{\sqrt{6}}I_3$, where $\lambda_a$ are Gell-Mann matrices. The covariant part can be written as:
	\begin{align}
	\label{eq_Wtriplet}
	\mathcal{W}_{\mu}\equiv W^a T^a=\frac{1}{2}\begin{pmatrix}
	W^3_{\mu} + \frac{1}{\sqrt{3}}W^8_{\mu}& \sqrt{2}W'^+_{\mu}  &\sqrt{ 2} Y'^+_{\mu}\\ 
	\sqrt{2} W^-_{\mu}& -W^3_{\mu}  + \frac{1}{\sqrt{3}}W^8_{\mu} &  \sqrt{2}V'^0_{\mu}\\ 
	\sqrt{2} Y^-_{\mu}& \sqrt{2} V'^{0*}_{\mu}&  - \frac{2}{\sqrt{3}}W^8_{\mu}
	\end{pmatrix},
	\end{align}
where we have defined   the mass eigenstates of the charged gauge bosons as  
\bea 
W'^{\pm}_{\mu}=\frac{1}{\sqrt{2}}\left( W^1_{\mu}\mp i W^2_{\mu}\right),\quad\,
Y'^{\pm }_{\mu}=\frac{1}{\sqrt{2}}\left( W^4_{\mu}\mp i W^5_{\mu}\right), \quad  V'^{0}_{\mu}=\frac{1}{\sqrt{2}}\left( W^6_{\mu}- i W^7_{\mu}\right).
\label{gbos}
\eea
\item For an $SU(3)_L$ antitriplet: $T^a=-\frac{1}{2}\lambda^*_a=-\frac{1}{2}\lambda^T_a~\forall a=1,2,..,8$, $T^9=\frac{1}{\sqrt{6}}I_3$. 
\item For an $SU(3)_L$ sextet denoted as $S\sim (6,2/3)$, given in table~\ref{tab:Fields_of_new_model}, the action of an $SU(3)_L$ generator can be written in terms of the  Gell-Mann matrix, $T^aS=S\lambda_a/2 +\lambda_a/2S^T $~\cite{Buras:2012dp}.  Hence, the corresponding covariant derivative can be written in terms of the generators of the $SU(3)$ triplet~\cite{Buras:2012dp,Pleitez:1997vx}, namely 
	\begin{equation}\label{eq_DS}
	D_{\mu}S=\partial_{\mu}S-ig\left[SW_{\mu} + SW_{\mu}^T\right] -ig_X\frac{ X}{\sqrt{6}}X_{\mu}S.
	\end{equation}
\end{itemize} 
The symmetry-breaking pattern is $SU(3)_L\times U(1)_X\overset{\langle\sigma^0_{i}\rangle}{\longrightarrow} SU(2)_L\times U(1)_Y\overset{\langle H^0_{\alpha}\rangle}{\longrightarrow}U(1)_Q$, where $i=1,2,S$ and $\alpha=1,2,3,S$. 

The covariant kinetic terms of the Higgs bosons are 
\begin{align}
\label{eq_lkHiggs}
L_{kin}^{H}=\sum_{i=1}^3\left(D_{\mu}\phi_i\right)^{\dagger}\left(D^{\mu}\phi_i \right) + \left(D_{\mu} S\right)^{\dagger}\left(D^{\mu}S \right).
\end{align}
From this, the squared mass matrix  of the charged gauge bosons  in the basis $(W'^{\pm}_{\mu},Y'^{\pm}_{\mu})$ is  given by
\begin{align}
\label{eq_chargedGmass1}
M^{2}_{V^{\pm}}=\frac{g^2}{2} \left(
\begin{array}{cc}
k_1^2+k_2^2+k_3^2+k_S^2+2 \varepsilon _S^2 & k_1 n_1+k_2 n_2+\sqrt{2} k_S n_S+\sqrt{2} k_S \varepsilon _S \\
k_1 n_1+k_2 n_2+\sqrt{2} k_S n_S+\sqrt{2} k_S \varepsilon _S & k_3^2+k_S^2+n_1^2+n_2^2+2 n_S^2 \\
\end{array}
\right).
\end{align} 
It is enough to assume that $k_{i}/n_i\ll 1$ for $i=2,S$ so that the non-diagonal term in the squared mass matrix~\eqref{eq_chargedGmass1} can be ignored. In  this work we will accept that 
\begin{align}
\label{eq_knratio}
 n_1=0, \quad \frac{k_2}{n_2}=\frac{k_s}{n_S}\ll1.
\end{align}
 In particular, we will choose  $k_{1,2,S}\sim \mathcal{O}(10)$ GeV and $n_{2,s}\sim \mathcal{O}(10^3)$ GeV, leading to the consequence that $k_in_i \mathrm{GeV^2}/(246 \mathrm{GeV})^2\ll1$. The non-zero values of $k_1$ still allow the reasonable Yukawa couplings of normal charged leptons given in Lagrangian~\eqref{eq_SMchargedlep}. We note that this choice of VEV values are still allowed for generating consistent quark masses,  as discussed previously~\cite{Fonseca:2016tbn}. The masses and physical states $\{W^{\pm},\, Y^{\pm}\}$ of  charged gauge bosons are determined as
\begin{align} \label{eq_physicalcgauge}
W^{\pm}&\simeq W'^{\pm},\quad m_W^2=\frac{g^2}{2} v^2,\quad v^2\equiv \left(k_1^2 +k_2^2 +k_3^2+k_S^2\right), \crn
Y^{\pm}&\simeq Y'^{\pm},\quad m_Y^2=\frac{g^2}{2} u^2, \quad u^2\equiv  \left(k_3^2 +k_S^2 + n_1^2 +n_2^2 +n_S^2\right). 
\end{align}
Identifying the $W^{\pm}$ with the SM one, we have $ v\simeq 174\,\mathrm{GeV}$.  If $k_{1,2,S}=\mathcal{O}(10) \,\mathrm{GeV}$, we have $k_3\simeq\, v$.  Using  the assumption in Eq.~\eqref{eq_knratio} the neutral gauge boson mass can be determined  as follows. 

The non-Hermitian gauge bosons $V^0$ and $V^{0*}$ do  not mix with the Hermitian ones. The masses and physical states are 
\begin{align} \label{eq_physicalNHgauge}
V^{0}&\simeq V'^{0},\quad m_{V}^2 = \frac{g^2}{2} \left(u^2 + n_S^2\right).
\end{align}
For simplicity in calculating  the masses and mass eigenstates of the Hermitian neutral gauge bosons, we will safely  use the limit that $k_1,k_2,k_S,\epsilon_S\ll k_3$. Accordingly, these neutral gauge bosons will decouple with the Re$V^0$.   In the basis $(X_{\mu}, W^3_{\mu}, W^8_{\mu})$, the squared mass matrix is 
\begin{align}
\mathcal{M}^2_{X38}=\frac{g^2}{2} \left(
\begin{array}{ccc}
\frac{2}{27} t^2 \left(3 n_S^2+u^2+4 v^2\right) & -\frac{2}{3} \sqrt{\frac{2}{3}} t v^2 & -\frac{2}{9} \sqrt{2} t \left(3
n_S^2+u^2+v^2\right) \\
-\frac{2}{3} \sqrt{\frac{2}{3}} t v^2 & v^2 & \frac{v^2}{\sqrt{3}} \\
-\frac{2}{9} \sqrt{2} t \left(3 n_S^2+u^2+v^2\right) & \frac{v^2}{\sqrt{3}} & \frac{1}{3} \left(12 n_S^2+4 u^2+v^2\right) \\
\end{array}
\right),
\end{align}
where $t=g_X/g$. This matrix will be diagonalized by a mixing matrix $C$ defined by
\begin{align}
\label{eq_M2x38diag}
M^2_{d}=C^T\mathcal{M}^2_{X38}C=M^2_{d}= \mathrm{diag}(0, M^2_{Z_1},M^2_{Z_2}). 
\end{align}
 This mixing matrix $C$  can be summarized in the three breaking steps as follows:  $ X_{\mu},\, W^3_{\mu}, \, W^8_{\mu} \xrightarrow{\theta_{331}}\, B_{\mu},\, W^3_{\mu}, \, Z'_{\mu}  \xrightarrow{\theta_W}  A_{\mu},\, Z_{\mu}, \, Z'_{\mu} \xrightarrow{\theta}   A_{\mu},\, Z_{1\mu}, \, Z_{2\mu}$ corresponding to three physical gauge bosons. Two of them are identified with  the massless photon $A_{\mu}$ and the SM-like neutral gauge boson $Z_1$ found experimentally.   After the first breaking step, the gauge couplings and $U(1)_Y$ charges  are identified  with the SM, leading to the following consequences:
\begin{align}\label{eq_SMmatching}
Y =\frac{1}{\sqrt{3}}T^8 +X,\quad 
t=\frac{g_X}{g} = \frac{3\sqrt{2}s_W}{\sqrt{3-4s^2_W}},
\end{align}
where $g$ and $s_W$ are the well-known parameters defined in the SM, i.e., the $SU(2)_L$ gauge couplings and the sine of the Weinberg angle.  
In the first step, the two neutral gauge bosons $W_\mu^8$ and $X_\mu$ mix, giving rise to the two bosons $B_\mu$ and $Z'_\mu$. The mixing angle is denoted by $\theta_{331}$ and is given by  \cite{Buras:2012dp}
\begin{align} \label{sc331}
s_{331}\equiv \sin\theta_{331}&=\frac{\sqrt{6}g}{\sqrt{6g^2+g_X^2/3}}=\sqrt{1-t_W^2/3}, \; 
c_{331}\equiv \cos\theta_{331} = \frac{t_W}{\sqrt{3}}.
\end{align}
 The relation between the original and physical basis of the  neutral gauge bosons are
 \begin{align}
 \begin{pmatrix}
 X_\mu \\
 W^3_\mu\\
 W^8_\mu
 \end{pmatrix} &= \begin{pmatrix}
 s_{331} &0 & c_{331}  \\
 0 & 1 & 0 \\
 c_{331} & 0 & -s_{331}
 \end{pmatrix}
 \begin{pmatrix}
 c_W & -s_W & 0 \\
 s_W & c_W& 0 \\
 0 & 0 & 1
 \end{pmatrix}
 \begin{pmatrix}
 1 &   0 & 0\\
 0 & c_\theta & -s_\theta \\
 0 & s_\theta & c_\theta
 \end{pmatrix}\begin{pmatrix}
 A_\mu \\
 Z_{1\mu} \\
 Z_{2\mu}
 \end{pmatrix}=C\begin{pmatrix}
 A_\mu \\
 Z_{1\mu} \\
 Z_{2\mu}
 \end{pmatrix},\crn 
 C&=\begin{pmatrix}
 s_{331}c_W, &   \left(- s_{331}s_Wc_\theta+c_{331}s_\theta\right),  & \left( s_{331}s_Ws_\theta +c_{331}c_\theta\right) \\
 s_W ,& c_Wc_\theta ,& -s_\theta c_w \\
 c_{331}c_W, &   -\left( c_{331}s_Wc_\theta+s_{331}s_\theta\right),  &\left(  c_{331}s_Ws_\theta -s_{331}c_\theta\right) 
 \end{pmatrix},
 \label{neutralgaugebosonmix}
 \end{align}
Using the limit $\epsilon_S^2\ll k_\al^2 \ll n_{2,S}^2$, the mixing angle $\theta$ is determined as \cite{Buras:2014yna}
\begin{align} \label{eq_stheta}
s_{\theta}\equiv \sin \theta & \simeq  \frac{\sqrt{3-4 s_W^2} v^2}{4 c_W^4 \left(u^2 +3 n_S^2\right)+2 \left(2 s_W^2-1\right) v^2}.
\end{align}
The masses for the neutral gauge bosons in this limit are 
\bea  \label{eq_mgauge}
m_A^2&=&0, \quad  m^2_{Z_1}\simeq\,m_Z^2= \frac{g^2v^2}{2c^2_W}, \quad  m^2_{Z_2}\simeq 
m_{Z'}^2 =  \frac{4g^2  c_W^2(u^2 +3n_S^2)}{3-4s_W^2}. 
\eea
As usual for 3-3-1 models with non-zero $Z-Z'$ mixing, in the limit $m^2_{Z'}\gg m^2_{Z}$  the tree-level contribution to the $\rho$ parameter defined by $\rho\equiv m^2_W/(m_Z^2c^2_W)$  is estimated approximately by the following formula~\cite{Buras:2014yna}
\begin{align}
\label{eq_drho}\Delta\rho\simeq \left(\frac{m^2_{Z'}}{m^2_Z}\right)s^2_{\theta}\simeq  \left(\frac{m^2_{Z}}{m^2_{Z'}}\right) \times \frac{2}{\sqrt{3-4 s_W^2} },
\end{align}
where $s_{\theta}$ is given in Eq.~\eqref{eq_stheta}.  The recent experimental lower bound of $m_{Z'}\geq 4$ TeV~\cite{Aaboud:2017sjh} results in that $\Delta\rho\leq 7\times 10^{-4}$, which still satisfies  $3\sigma$ allowed range of experimental data~\cite{Tanabashi:2018oca}. Previous  studies of one-loop contributions from heavy gauge and Higgs bosons to the $\rho$ parameter in some particular 3-3-1 models~\cite{Hoang:1999yv, Buras:2014yna, Long:2018dun} suggest that these contributions from the heavy gauge bosons are very suppressed with $m_{Z'}\ge 4$ TeV, while those from  Higgs bosons  can be negative and have the order of $\mathcal{O}(10^{-4})$.  Hence the total contributions to $\Delta\rho$  may satisfy the experimental constraint even with  $m_{Z'}$ smaller than 4 TeV, which was reported from the ATLAS  experiment at LHC~\cite{Aaboud:2017sjh}.  We will use this lower bound of $m_{Z'}$ in the numerical investigation.

To determine the SM-like Higgs from its couplings to the gauge bosons $W^{\pm}$ and $Z$, the relevant terms are  
 \begin{align}
 \label{eq_Lgh0}
 \mathcal{L}_{VS} =& \frac{g^2}{2}(W^+.W^-)  \left[ \sum_{i=1}^3 2k_{i} R_{i} +2 k_S R_{S} + 4 \epsilon_S R_{\Delta} +\sum_{i=1}^3 R^2_{i} + R^2_{S} + 2 R_{\Delta}^2\right]  \crn 
 &+ \frac{g^2}{4 c_W^2}  Z^2  \left[ \sum_{i=1}^3 2k_{i} R_{i} +2 k_S R_{S} + 8 \epsilon_S R_{\Delta} +\sum_{i=1}^3 R^2_{i} + R^2_{S} + 4 R_{\Delta}^2\right]. 
 \end{align}
In the limit $k_{1,2,S},\epsilon_S\ll k_3$, we have $k_3\simeq v=\sqrt{2}m_W/g$. Then we can see that $R_3$ should be identified with the SM-like Higgs boson because it has the same couplings with the SM gauge bosons as those predicted by the SM. 

As noted in previous works,   $m^2_{Z_2}\gg m^2_Z$,  so we get $s_{\theta}\ll1$ based on Eq.~\eqref{eq_stheta}, hence the $Z-Z'$ mixing will be ignored in one-loop formulas  involving with LFV decays.  An interesting property of the heavy gauge bosons is that they get masses from two large vev $n_2$ and $n_S$.  Hence, in principle, $n_2$ can get low values of 1 TeV, even when $m_{Z'}$ are constrained  to be very heavy from recent experiments.   

 \subsection{Higgs boson}
 The Higgs potential is~\footnote{We thank the referee for  pointing out a missing term of this Higgs potential in the previous version.}:
 \begin{align}\label{eq_Vh}
 V_h&=V(\phi_1,\phi_2,\phi_3)+ V(S) +V(S,\phi), \crn 
 V(\phi_1,\phi_2,\phi_3)&= \sum_{i=1}^3\left[ \mu_i^2 \phi^{\dagger}_i\phi_i + \lambda^{\phi}_i \left(\phi^{\dagger}_i\phi_i\right)^2\right] +\left( \mu_{12}^2\phi^{\dagger}_1\phi_2 +\mathrm{h.c.}\right) \crn 
 &+ \sum_{i<j,i,j=1}^3 \left[ \lambda^{\phi}_{ij} \left(\phi^{\dagger}_i\phi_i\right)\left(\phi^{\dagger}_j \phi_j\right) + \tilde{\lambda}^{\phi}_{ij} \left(\phi^{\dagger}_i\phi_j\right)\left(\phi^{\dagger}_j \phi_i\right)\right] + \left[ \lambda^{\prime \phi}_{12}\left(\phi^{\dagger}_1\phi_2\right)^2 +\mathrm{h.c.}\right] \crn 
 & - \sum_{i<j<k, i,j,k=1}^3\sqrt{2}f^{\phi}\left(\epsilon_{ijk}\phi_i \phi_j\phi_k +\mathrm{h.c.}\right),\crn
 V(S)&=\mathrm{Tr}\left[ \mu_S^2  (S^{\dagger}S)  +\lambda^S_1(S^{\dagger}S)^2  \right] + \lambda^S_{2}\left[\mathrm{Tr}(S^{\dagger}S) \right]^2, \crn 
 V(S,\phi)&= \mathrm{Tr}(S^{\dagger}S)  \sum_{i=1}^3 \lambda^{\phi S}_i \phi^{\dagger}_i\phi_i  + \mathrm{Tr}(S^{\dagger}S)  \left( \lambda^{\phi S}_{12} \phi^{\dagger}_1\phi_2 +\mathrm{h.c.} 
 \right)   \crn 
 & + \sum_{i=1}^3 \tilde{\lambda}^{\phi S}_i\left[  \phi^{\dagger}_iSS^*\phi_i\right] +   \left( \tilde{\lambda}^{\phi S}_{12} \phi^{\dagger}_1SS^*\phi_2 +\mathrm{H.c.} 
 \right)  \crn 
 &+\sum_{i<j,i,j=1}^2f^{\phi S}_{ij}\left( \phi^T_iS^*\phi_j +\mathrm{H.c.}\right) + \lambda'^{\phi S} \left[ (\phi^*_2)_{\ell}S_{\ell i}(\phi_1)_j(\phi_3)_k \epsilon_{ijk} +\mathrm{H.c.}\right],
 \end{align}
where the invariant terms containing Higgs sextets were derived based on ref.~\cite{Diaz:2003dk}, $\epsilon_{ijk}$ is the total antisymmetric tensor.

 For one-loop contributions of Higgs bosons to  LFV decays  of the  SM-like Higgs boson and charged leptons, we pay attention to Higgs components appearing in the Yukawa terms given in Eq.~\eqref{eq_SMchargedlep}. Furthermore, if $\Sigma^-$ or $E_e$ does not mix with $E_{\mu}$, the LFV decays containing $e$ as a final state are suppressed;  these get Higgs contributions from light active neutrinos and singly charged Higgs exchanges. Here, the simple case of $k_1\ll k_3$ allows us to take $k_1\simeq0$ in the squared mass matrices of all Higgs bosons.  We note that $k_1\neq 0$ is still necessary for generating right quark masses as well as couplings of the SM-like Higgs boson with normal charged leptons. 
 
 For simplicity in  finding physical states and masses of  neutral Higgs bosons, we use the following limit: 
 \begin{align} \label{eq_laLimit} 
 \lambda '^{\phi  S}\to 0,\lambda _{23}^{\phi }\to 0,\lambda _3^{\phi  S}\to 0,\tilde{\lambda }_{12}^{\phi  S}\to -2 \lambda _{12}^{\phi  S},\lambda _2^{\phi  S}\to -\frac{f_{22}^{\phi S}}{n_S}-\tilde{\lambda }_2^{\phi  S}. 
 \end{align}
 We remind  the reader of the other assumptions that we mentioned above that can be applied for finding physical states of the Higgs bosons:  $k_2,k_s,k_1\simeq 0$ and $n_1=0$. There are eight  neutral Higgs components in the Higgs sector, corresponding to eight equations of the minimum conditions of the Higgs potential.   The minimal equations  are listed in Appendix~\ref{app_minVhigg}. Inserting them into the Higgs potential~\eqref{eq_Vh}, we  will find the masses and mixing matrices of all physical Higgs bosons as follows.
 
 There are  six physical states of CP-even neutral Higgs bosons that  are the original states themselves, namely 
 \begin{equation}\label{eq_Hphysical}
 R_{1}\equiv h^0_1,\; R_3\equiv h,\; R_{\sigma_1}\equiv h^0_2, \; R_{\sigma_2}\equiv h^0_3,\; R_{\sigma_S}\equiv h^0_4,\; R_{\Delta}\equiv h^0_5
 \end{equation} with corresponding masses as follows:
 \begin{align}
 \label{eq_NevenHiggs}
 m^2_{R_{1}} &= \mu'^2_{1}=2 k_1^2 \lambda _1^{\phi }+k_3^2 \lambda _{13}^{\phi }+n_2^2 \lambda _{12}^{\phi }+n_S^2 \lambda _2^{\phi S}, \crn 
   m^2_{R_{3}}&=4 \lambda_3 k_3^2,\; m^2_{R_{\sigma_1}}=\mu'^2_{1} +\left( \lambda _{12}^{\phi } +2\lambda _{12}^{\prime \phi } \right)  n^2_2 +\tilde{\lambda }_1^{\phi S} n_S^2,\;  m^2_{\sigma_{2}}= 4\lambda_{2}n_2^2, \crn 
  m^2_{\sigma_{S}}&= 4 n_S^2 (\lambda _1^S+\lambda _2^S) -\frac{n_2^2 f_{22}^{\phi {S}}}{n_S},\; m^2_{\Delta}=-\frac{n_2^2 (n_S \tilde{\lambda }_2^{\phi S}+f_{22}^{\phi })}{n_S}-2 n_S^2 \lambda _1^S. 
 \end{align}
 The squared matrix of the two states $(R_{2},R_{S})$ is 
 \begin{align}
M^2_{2S}= \left(
 \begin{array}{cc}
 -n_S (n_S \tilde{\lambda }_2^{\phi S}+2 f_{22}^{\phi S}) & \frac{n_2 (n_S \tilde{\lambda }_2^{\phi S}+2 f_{22}^{\phi S})}{\sqrt{2}} \\
 \frac{n_2 (n_S \tilde{\lambda }_2^{\phi S}+2 f_{22}^{\phi S})}{\sqrt{2}} & -\frac{n_2^2 (n_S \tilde{\lambda }_2^{\phi S}+2 f_{22}^{\phi S})}{2 n_S} \\
 \end{array}
 \right),
 \end{align}
 which gives give two mass eigenstates corresponding to one Goldstone boson of $V^0$ and one physical state, which are denoted as $G_{V}$ and  $h^0_6$. Their masses and relations to the original states are 
 \begin{align} \label{eq_h06}
 m^2_{G_V}&=0,\; m^2_{h^0_6}= -\left(n_2^2 +2 n_S^2\right)\left(\frac{f_{22}^{\phi S}}{n_S}+\frac{\tilde{\lambda }_2^{\phi S}}{2}\right),  \crn  \begin{pmatrix}
 R_2\\ 
 R_S
 \end{pmatrix} &=
 \left(
 \begin{array}{cc}
 c_{2s} & -s_{2s} \\
 s_{2s} & c_{2s} \\
 \end{array}
 \right) \begin{pmatrix}
 G_{V}\\ 
 h^0_6
 \end{pmatrix}, \;c_{2s}=\frac{n_2}{\sqrt{n_2^2 +2n_S^2}},\; s_{2s}=\frac{\sqrt{2}n_S}{\sqrt{n_2^2 +2n_S^2}}. 
 \end{align}
 We can see that the above assumptions of the VEV and Higgs self-couplings gives one Goldstone boson $G_V$ of the non-Hermitian gauge boson $V$ and a light CP-even neutral Higgs boson $h\equiv R_3$. It will be identified with the SM-like Higgs boson found by LHC through its couplings with fermions and gauge bosons, as we will show later. 
 
  The model contains only one pair of doubly charged Higgs bosons  $\Delta^{\pm\pm}$ with mass 
 \begin{equation}\label{eq_ccH}
 m^2_{\Delta^{++}}= k_3^2 \tilde{\lambda }_3^{\phi S}+n_2^2 \left(-\frac{f_{22}^{\phi S}}{n_S}-\tilde{\lambda }_2^{\phi S}\right)-2 n_S^2 \lambda _1^S. 
 \end{equation}
 Regarding  singly charged scalars, we have found two zero mass eigenvalues corresponding to two Goldstone bosons of $W^{\pm}$ and $Y^{\pm}$.  There are three original states that are also the mass eigeinstates,
 \begin{align}
 \label{eq_1cHiggs}
 G^{\pm}_{W}&\equiv H^{\pm}_3, \; m_{G_W}=0, \crn 
 m^2_{H^{+}_1}&=\mu'^2_1+k_3^2 \tilde{\lambda }_{13}^{\phi }, \quad
  m^2_{\Delta^{+}}=\frac{1}{2} \left(k_3^2 \tilde{\lambda }_3^{\phi S}-\frac{2 n_2^2 (n_S \tilde{\lambda }_2^{\phi S}+f_{22}^{\phi S})}{n_S}-4 n_S^2 \lambda
  _1^S\right).
 \end{align}   
 Corresponding to three other singly charged Higgs states $(H^{\pm}_3,\sigma^{\pm},H^{\pm}_S)$, the squared mass matrix  is 
 \begin{align}
 M^2_{3\sigma S}=\left(
 \begin{array}{ccc}
 k_3^2 \tilde{\lambda }_{23}^{\phi }-n_S (2 f_{22}^{\phi S}+n_S \tilde{\lambda }_2^{\phi S}) & k_3 n_2 \tilde{\lambda }_{23}^{\phi } & \frac{n_2 (2
 	f_{22}^{\phi S}+n_S \tilde{\lambda }_2^{\phi S})}{\sqrt{2}} \\
 k_3 n_2 \tilde{\lambda }_{23}^{\phi } & \tilde{\lambda }_{23}^{\phi } n_2^2+n_S^2 \tilde{\lambda }_3^{\phi S} & \frac{k_3 n_S \tilde{\lambda
 	}_3^{\phi S}}{\sqrt{2}} \\
 \frac{n_2 (2 f_{22}^{\phi S}+n_S \tilde{\lambda }_2^{\phi S})}{\sqrt{2}} & \frac{k_3 n_S \tilde{\lambda }_3^{\phi S}}{\sqrt{2}} & \frac{1}{2}
 \left(k_3^2 \tilde{\lambda }_3^{\phi S}-\frac{n_2^2 (2 f_{22}^{\phi S}+n_S \tilde{\lambda }_2^{\phi S})}{n_S}\right) \\
 \end{array}
 \right).
 \end{align}
 It is easily seen that Det$[M^2_{3\sigma S}]=0$, leading to a massless eigenstate that can be identified with the Goldstone boson of $V^{\pm}$. 
 
 In the CP-odd neutral Higgs spectrum, there are three massless eigenstates corresponding to three Goldstone bosons of gauge bosons $Z,Z'$ and $V^0$. In particular, the three mass eigenstates and two Goldstone bosons  are  
 \begin{align}
 \label{eq_IHiggs}
 m^2_{I_1} &= \mu'^2_1, m^2_{I_{\sigma_1}}=\quad m^2_{I_{\Delta}}= -\frac{n_2^2 (n_S \tilde{\lambda }_2^{\phi S}+f_{22}^{\phi S})}{n_S}-2 n_S^2 \lambda _1^S, \crn 
   m^2_{G_Z}&=0, \; \quad G_Z\equiv I_3,
 \end{align}
 where $G_Z$ is the Goldstone boson absorbed by the gauge boson $Z$. Five remaining states divide into two sub-matrices of squared masses, corresponding to bases $(I_2, I_S)$ and $(I_{\sigma_1},I_{\sigma_2},I_{\sigma_S})$, namely 
 \begin{align}
 \label{eq_I5mass}
 M^2_{I2s}&= \left(
 \begin{array}{cc}
 -n_S (2 f_{22}^{\phi S}+n_S \tilde{\lambda }_2^{\phi S}) & \frac{n_2 (2 f_{22}^{\phi S}+n_S \tilde{\lambda }_2^{\phi S})}{\sqrt{2}} \\
 \frac{n_2 (2 f_{22}^{\phi S}+n_S \tilde{\lambda }_2^{\phi S})}{\sqrt{2}} & -\frac{n_2^2 (2 f_{22}^{\phi S}+n_S \tilde{\lambda }_2^{\phi
 		S})}{2 n_S} \\
 \end{array}
 \right), \crn 
  M^2_{\sigma_{1,2}\Delta} &=\left(
  \begin{array}{ccc}
 \left(\tilde{\lambda }_{12}^{\phi } -2\tilde{\lambda }_{12}^{\prime\phi }\right) n_2^2+n_S^2 \tilde{\lambda }_1^{\phi S}+\mu'^2_1 & -4 n_S^2 \lambda _{12}^{\phi S} & 2 n_2 n_S \lambda
  _{12}^{\phi S} \\
  -4 n_S^2 \lambda_{12}^{\phi S} & -4 n_S f_{22}^{\phi S} & 2 n_2 f_{22}^{\phi S} \\
  2 n_2 n_S \lambda_{12}^{\phi S} & 2 n_2 f_{22}^{\phi S} & -\frac{n_2^2 f_{22}^{\phi S}}{n_S} \\
  \end{array}
  \right).
 \end{align}
 The first $2\times2$ matrix gives one Goldstone boson of $V^0$ denoted as $ G'_{V}$, $m_{G'_V}=0$, and a physical CP-odd neutral Higgs  $a_6$. Their mass and mixing matrix is 
 \begin{align}
 \label{eq_I2s}
 m_{G'_V}&=0,\quad m^2_{a_6}=\left(-n_2^2-2 n_S^2\right) \left(\frac{f_{22}^{\phi S}}{n_S}+\frac{\tilde{\lambda }_2^{\phi S}}{2}\right),\crn 
 \begin{pmatrix}
 I_2\\ 
 I_S
 \end{pmatrix}&=  \left(
 \begin{array}{cc}
 c_{2s} & -s_{2s} \\
 s_{2s} & c_{2s} \\
 \end{array}
 \right)  \begin{pmatrix}
 G'_V\\ 
 a_6
 \end{pmatrix}. 
 \end{align}
 Regarding to the second matrix in Eq.~\eqref{eq_I5mass},  it is easy to check that Det$[M^2_{\sigma_{1,2}\Delta}]=0$;  equivalently, there exists one massless state that can be identified with the Golstone boson of $Z'$. Because $I_{\sigma_2}$ and $I_{\Delta}$ are irrelevant with the couplings in Eq.~\eqref{eq_SMchargedlep}, which contribute to the one-loop amplitude of  LFV decays, we choose a simple case that $\lambda_{12}^{\phi S} =0$ so that $I_{\sigma_1}$ is itself physical.   The CP-odd neutral Higgs bosons relating to the one-loop contributions to LFV decays are $I_{\sigma_1}$ and $a_6$. 
 
 According to the above discussion on the Higgs sector, we can see that $R_{\sigma_1}$ and  $I_{\sigma_1}$ are the real and imaginary parts of a physical Higgs boson $\sigma_1$ with mass $m^2_{\sigma_1}=\left(\tilde{\lambda }_{12}^{\phi } -2\tilde{\lambda }_{12}^{\prime\phi }\right) n_2^2 +\tilde{\lambda }_1^{\phi S}
 n_S^2 +\mu'^2_1$.  Similarly, there is another neutral complex Higgs boson denoted as $h_6=(h^0_6 +i a_6)/\sqrt{2}$ with mass $m^2_{h_6}=m^2_{a_6}=m^2_{h^0_6}$ given in Eqs.~\eqref{eq_h06} and \eqref{eq_I2s}. 
 
 According to the above discussion on the Higgs sector,  we can see that $h^0_6$ and  $a_6$ can be considered as real and imaginary parts of a physical neutral complex Higgs boson denoted as $h_6\equiv (h^0_6 +i a_6)/\sqrt{2}$   $\sigma_1$  with squared mass $m^2_{h_6}=m^2_{a_6}=m^2_{h^0_6}$ given in Eqs.~\eqref{eq_h06} and \eqref{eq_I2s}.  Similarly, in the limit of the unknown parameter $\tilde{\lambda }_{12}^{\prime\phi } =0$, $R_{\sigma_1}$ and  $I_{\sigma_1}$  can be considered as  the real and imaginary parts of a physical Higgs boson $\sigma_1$ with mass $m^2_{\sigma_1}= \tilde{\lambda }_{12}^{\phi } n_2^2 +\tilde{\lambda }_1^{\phi S}n_S^2 +\mu'^2_1$. More interesting,  $R_{\sigma_1}$ and  $I_{\sigma_1}$ give the same qualitative contributions to the  amplitudes  of the LFV  decays. Therefore, we will use this limit  for our numerical investigation to avoid unnecessary and lengthy private one-loop contributions of $R_{\sigma_1}$ and  $I_{\sigma_1}$ to LFV decay amplitudes.
 
From  the simple Higgs potential shown above, the  Feynman rules for Higgs self-couplings of the SM-like Higgs boson that contribute to the LFVHD are shown in Table~\ref{table_hhh}. Note that the coupling $h h_6h_6$ is zero. 
\begin{table}[ht]
	\centering 
	\begin{tabular}{cccc}
		\hline 
		Coupling &  Vertex & Coupling &  Vertex\\ 
		\hline 
		$h \sigma^0_1\sigma^{0*}_1$	& $-i\lambda_{13}m_W/g$ &	$h\sigma^0_1h_6$		& $if^{\phi} s_{2s}/2$  \\  
		\hline 
	\end{tabular} 
	\caption{Feynman rules for Higgs self-couplings that contribute to LFVHD  decays.}\label{table_hhh}
\end{table}
After determining the masses and mixing matrices of all leptons, gauge and Higgs bosons,  the branching ratios of LFV decays $h\rightarrow e_be_a$ and $e_b\rightarrow e_a\gamma$ can be computed in the next section. 

\section{\label{sec_LFVdecay} LFV decays $e_b\rightarrow e_a \gamma$ and $h\rightarrow e_ae_b$ }
\subsection{Analytic formulas of branching ratios}
In this section, we only pay attention to  couplings that contribute to the LFV decay amplitudes  $h\rightarrow e_be_a$ and $e_b\rightarrow\,e_a\gamma$ at the one-loop level.  We also apply the results introduced in Ref.~\cite{Hue:2015fbb} to calculate the amplitudes of the decays $h\rightarrow e_ae_b$. In this model, couplings of charged leptons with active neutrinos  result in suppressed contributions to the LFV decay, similar to the case of the SM  with very light neutrinos. Hence the  non-trivial LFV couplings with normal charged  leptons that give large LFV effects  relate to only heavy charged leoptons $E_i$, leading to that the LFV couplings that we consider here being only $V^0 \overline{E_{i}}e_a$ or $s^0 \overline{E_{i}}e_a$, and their Dirac conjugations. 

 The $\overline{f}fV$ couplings  are contained in the covariant kinetic terms of leptons,
 \begin{align}
 \mathcal{L}_{ffV}=\sum_{i=1}^7 i\overline{e_{iR}}\gamma^{\mu}D_{\mu} e_{iR} + \sum_{i=e,\mu,\tau} i\overline{L_{i}}\gamma^{\mu}D_{\mu} L_{i},
 \end{align}
  see  the detailed explanation of the relations between these notations in Ref.~\cite{Dreiner:2008tw}. The following terms are involved with LFV couplings:
  \begin{align} \label{eq_LFVcoup}
 \mathcal{L}^{\mathrm{LFV}}_{ffV}&= g \left[ \overline{E_{eL}} \gamma^{\mu}e_L +\frac{1}{\sqrt{2}} \left( \overline{E_{\mu L}} \gamma^{\mu}\mu_L + \overline{E_{\tau L}} \gamma^{\mu}\tau_L \right)\right] V^0_{\mu} + \mathrm{h.c.} \crn
 &=g \left[ (V^{E*}_L)_{3i}\overline{E_{i}} \gamma^{\mu}P_Le +\frac{1}{\sqrt{2}} \left[ (V^{E*}_L)_{1i}\overline{E_{i}} \gamma^{\mu}P_L\mu + (V^{E*}_L)_{2i}\overline{E_{i}} \gamma^{\mu}P_L \tau \right]\right] V^0_{\mu} + \mathrm{h.c.}
 \end{align}
 
 Based on the general Feynman rules for one-loop contributions to the decay amplitude $h\rightarrow e_ae_b$, the diagrams need  vertices  with non-zero couplings $hV^0V^{0*}$, or $hs^0V^{0*}$, where $s^0$ is a neutral Higgs boson. In the model under consideration these kinds of couplings do not appear in the model.  In contrast, the couplings given in Eq. \eqref{eq_LFVcoup} do contribute to the decay amplitudes $e_b\rightarrow e_a\gamma$. 
 
 The $\overline{f}fs^0$ couplings come from the Yukawa Lagrangian~\eqref{eq_SMchargedlep}. In the physical basis, the Yukawa couplings  involved to LFVHD are
 \begin{align}
 \label{eq_h0ff}
 \mathcal{L}_{s^0ff}&=  -\frac{H^{0*}_1}{k_1}  \left[m_{\mu} \overline{\mu_R} \mu_L +m_{\tau} \overline{\tau_R} \tau_L \right] -  \sigma^{0}_1 \sum_{i=1}^3 \left[\frac{m_{\mu}}{k_1}  \left(V^{E*}_{L}\right)_{1i}\overline{E_i}P_R \mu + \frac{m_{\tau}}{k_1} \left(V^{E*}_{L}\right)_{2i}\overline{E_i}P_R \tau  \right]   \crn 
& - h^*_6\sum_{i=1}^3\left[ s_{2s} \left(Y^{\ell}_{1i} \overline{E_{i}}  P_L\mu + Y^{\ell}_{2i}\overline{E_{i}}P_L\tau\right) +  \frac{c_{2s} n_S}{n_2}Y^{\ell}_{3i} \overline{E_{i}}P_Le\right]  +\mathrm{h.c.},
  \end{align}
 where the matrix $Y^{\ell}$ is given in Eq.~\eqref{eq_massEmutaue}, which can be written in terms of heavy charged lepton masses and mixing parameters  based on Eq.~\eqref{eq_E123}: 
 \begin{align}
 \label{eq_Yiab}
 Y^{\ell}= \frac{1}{n_2} \mathrm{diag}(m_{E_1},\,m_{E_2},\, m_{E_3})V^{E\dagger}_L,
 \end{align}
  For convenience in calculating the one-loop contributions of  Higgs mediation to the LFV amplitudes,
    Lagrangian~\eqref{eq_h0ff} is written in the following  form: 
  \begin{align}
 \label{eq_h0ffR}
 \mathcal{L}_{sff}&= -\frac{H^{0*}_1}{k_1}  \left[m_{\mu} \overline{\mu_R} \mu_L +m_{\tau} \overline{\tau_R} \tau_L \right]  - \sigma^{0}_1\sum_{i=1}^3\sum_{j=1,2}Y^{\sigma^0_1}_{ji}\overline{E_i} P_R e_{(j+1)}   \crn 
 &- h^*_6\sum_{i}^3\left[ \sum_{j=1}^2 Y^{h_6}_{ji} \overline{E_i} P_L e_{(j+1)} +Y^{h_6}_{3i} \overline{E_i} P_L e \right]+\mathrm{h.c.}. 
 \end{align}
 where the coupling $Y^s_{ji}$, $i,j=1,2,3$,  is defined as follows: 
 \begin{align} \label{eq_Ys}
Y^{\sigma^0_1}_{ji}= \left\{ \begin{array}{cc}
\frac{m_{e_{(j+1)}}}{k_1}(V^{E*}_L)_{ji}, &  \;j=1,2,  \\ 
0,  & j=3
 \end{array} \right.,  \quad 
  Y^{h_6}_{ji}=     \left\{ \begin{array}{cc}
  s_{2s}Y^{\ell}_{ji};, &  \;j=1,2,  \\ 
  \frac{s_{2s} }{\sqrt{2}}Y^{\ell}_{ji},  & j=3
  \end{array} \right.,   
 \end{align}
 where we have used $s_{2s}=\frac{c_{2s}\sqrt{2} n_S}{ n_2}$.

\begin{figure}[ht]
	\includegraphics[width=16cm]{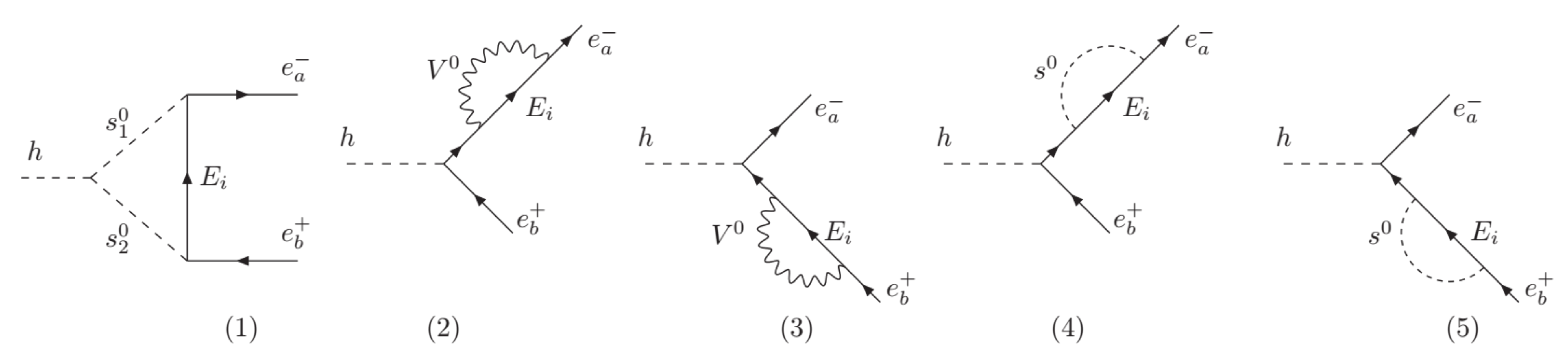}
	\caption{One-loop Feynman diagrams for decay $h\rightarrow{e}_a e_b$ in the unitary gauge, where $s^0,s^0_1,s^0_2=h_6,\sigma^0_1$.} \label{fig_SMlikeHtoeiej}
\end{figure}

The corresponding one-loop Feynman diagrams that contribute to the LFVHD amplitude are shown in Fig.~\ref{fig_SMlikeHtoeiej}. 
Although the model under consideration contains charged Higgs bosons, their one-loop contributions to the LFV decay are tiny.  The LFV couplings of the  doubly and singly  charged Higgs  bosons $\Delta^{\pm\pm}$ and $H^{\pm}_S$ do not appear because they only couple with electron; see Eq.~\eqref{eq_SMchargedlep}. The other singly charged Higgs bosons only couple with  active neutrinos having tiny masses;  hence one-loop contributions  involving with them  to LFV decay amplitudes  are proportional to the deviations between the squared masses of  the active neutrinos $\Delta m^2_{ij}\equiv m^2_i-m^2_j$, with $i\neq j$ and $i,j=1,2,3$. This result can be derived  using Taylor expansion in terms of  the squared masses of the active neutrinos and applying the Glashow-Iliopoulos-Maiani (GIM) mechanism $\sum_{i} V^*_{ia}V_{ib}=0$ to cancel large contributions  independent of $m_i$, see  previous discusion on LFV decays ~\cite{Cheng:1985bj, Thuc:2016qva}. Hence these contributions from singly charged Higgs bosons are very suppressed so  we then safely ignore them.

The partial decay width of the decays $h\rightarrow\,e_ae_b$ is defined as follows:
\be
\Gamma (h\rightarrow\,e_a e_b)\equiv\Gamma (h\rightarrow e_a^{-} e_b^{+})+\Gamma (h\rightarrow e_a^{+} e_b^{-})
=  \fr{ m_{h} }{8\pi }\left(\vert \Delta_{(ba)L}\vert^2+\vert \Delta_{(ba)R}\vert^2\right), \label{LFVwidth} 
\ee
with the condition  $m_{h}\gg m_{a,b}$ and $m_{a,b}$ charged lepton, $a,b=1,2,3$ corresponding to $e,\mu,\tau$. The on-shell conditions for external particles are $p^2_{a,b}=m_{a,b}^2$ and $ p_{h}^2 \equiv( p_a+p_b)^2=m^2_{h}$. The  LFVHD decay rate  is  Br$(h\rightarrow e_ae_b)= \Gamma (h\rightarrow \,e_ae_b)/\Gamma^{\mathrm{total}}_{h}$ where $\Gamma^{\mathrm{total}}_{h}=4.1\times 10^{-3}$ GeV. In the notations constructed in Ref.~\cite{Hue:2015fbb}, the $\Delta_{(ba)L,R}$ can be written as
\be \Delta_{(ba)L,R} =\sum^{5}_{i=1} \Delta^{(i)}_{(ba)L,R},  \label{deLR}\ee
where   detailed calculations to derive analytic formulas of $\Delta^{(i)}_{(ba)L,R}$  are given in Ref.~\cite{Hue:2015fbb}.   In previous works \cite{Thao:2017qtn,Hue:2015fbb}, we can see that $\Delta^{(2+3)}_{(ba)L,R}$ and $\Delta^{(4+5)}_{(ba)L,R}$ are very suppressed, hence we focus only to $\Delta^{(1)}_{(ba)L,R}=\Delta_{(ba)L,R}$ with the following analytic forms for non-zero contributions: 
\begin{align}
\label{eq_Deltaba}
\Delta_{(32)L,R}&=\Delta^{\sigma^0_1\sigma^0_1}_{(32)L,R} +\Delta^{\sigma^0_1h_6}_{(32)L,R} + \Delta^{h_6\sigma^0_1}_{(32)L,R}, \crn 
\Delta_{(b1)L,R}&=\Delta^{\sigma^0_1h_6}_{(b1)L,R}, 
\end{align}
where  $b=2,3$, and 
\begin{align}\label{eq_Deltaij}
\Delta^{\sigma^0_1\sigma^0_1}_{(32)L}=& \frac{m_{\tau}\lambda_{13}m_W}{16\pi^2g}\times  \sum_{i=1}^3Y^{\sigma^0_1*}_{1i}Y^{\sigma^0_1}_{2i} \left[-C_2 (0,0; m^2_{E_i},m^2_{\sigma^0_1},m^2_{\sigma^0_1})\right], \crn
\Delta^{\sigma^0_1\sigma^0_1}_{(32)R}=&  \frac{m_{\mu}\lambda_{13}m_W}{16\pi^2g}\times  \sum_{i=1}^3Y^{\sigma^0_1*}_{1i}Y^{\sigma^0_1}_{2i}  \left[C_1(0,0; m^2_{E_i},m^2_{\sigma^0_1},m^2_{\sigma^0_1})\right], \crn
\Delta^{\sigma^0_1h_6}_{(32)L}=& -\frac{f^{\phi}s_{2s}}{32\pi^2}\times  \sum_{i=1}^3Y^{\sigma^0_1*}_{1i}Y^{h_6}_{2i}  \left[m_{E_i}C_0(0,0; m^2_{E_i},m^2_{\sigma^0_1},m^2_{h_6})\right], \crn
\Delta^{\sigma^0_1h_6}_{(32)R}=& 0, \crn 
\Delta^{h_6\sigma^0_1}_{(32)L}=& 0, \crn
\Delta^{h_6\sigma^0_1}_{(32)R}=& -\frac{f^{\phi}s_{2s}}{32\pi^2}\times  \sum_{i=1}^3Y^{\sigma^0_1}_{2i}Y^{h_6*}_{1i}  \left[m_{E_i}C_0(0,0; m^2_{E_i},m^2_{h_6}, m^2_{\sigma^0_1})\right], \crn
\Delta^{\sigma^0_1h_6}_{(b1)L}=& 0, \crn
\Delta^{\sigma^0_1h_6}_{(b1)R}=& -\frac{f^{\phi}s_{2s}}{32\pi^2}\times  \sum_{i=1}^3Y^{\sigma^0_1}_{3i}Y^{h_6*}_{(b-1)i}  \left[m_{E_i}C_0(0,0; m^2_{E_i},m^2_{h_6}, m^2_{\sigma^0_1})\right]. \nn 
\end{align}
The functions $ C_{1,2}(0,0;m^2_{E_i},m^2_{s^0_1},m^2_{s^0_2})\equiv C_{0,1,2}(m_{E_i},m_{s^0_1},m_{s^0_2})$  are one-loop three-point Passarino-Veltman (PV) functions introduced in Ref.~\cite{Hue:2015fbb}. 

The $\Delta^{\sigma^0_1h_6}_{(32)L}$ arises from the chirality flip in the Yukawa couplings of heavy fermions with $\sigma^0_1$ and $h^0_6$ given in Eq.~\eqref{eq_h0ffR},  similar to the cases mentioned in Refs.~\cite{Baek:2015fma, Herrero-Garcia:2016uab}, which relates to the Yukawa couplings with chirality flip. In our work, the $\Delta^{\sigma^0_1h_6}_{(32)L,R}$ arises from the chirality flip in the Yukawa couplings of heavy fermions with $\sigma^0_1$ and $h^0_6$ given in Eq.~\eqref{eq_h0ffR}. This may give an interesting result that Br$(h\rightarrow e_be_a)$ may be large with large Yukawa couplings of $E_i$ in the perturbative limit. 

\begin{figure}[ht]
\includegraphics[width=12cm]{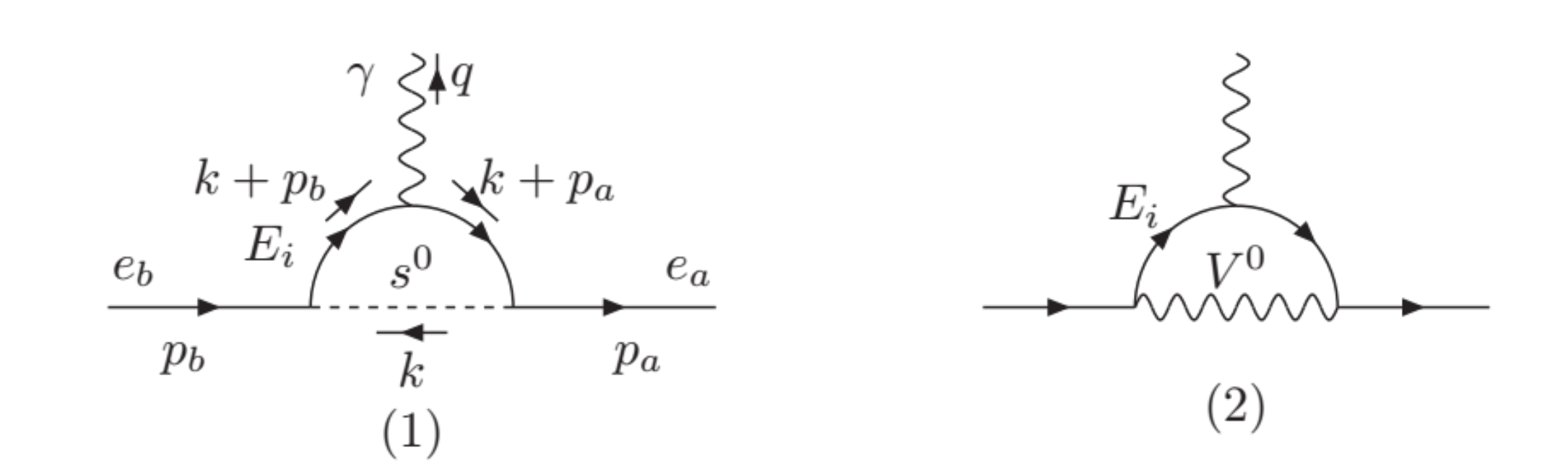}
	\caption{One-loop diagrams contributing to $C_{(ba)L,R}$ for decays ${e}_b\rightarrow {e_a}\gamma$, where $s^0=\sigma^0_1,h_6$. } \label{fig_eitoejgamma}
\end{figure}

In the unitary gauge, the one-loop three-point Feynman diagrams contributing to the decay amplitudes $e_b \rightarrow{e_a}\gamma$ ($a<b$) are shown in Fig.~\ref{fig_eitoejgamma}. 

For low energy,  the branching ratios of the cLFV decays can be written in a more convinient form  as follows:   
\begin{equation}\mathrm{Br}(e_b\rightarrow e_a\gamma)= \left(1- \frac{m_a^2}{m_b^2}\right)^3 \times \frac{3\alpha_{\mathrm{e}}}{2\pi}\left(\left| F_{(ba)L} \right|^2+ |F_{(ba)R}|^2\right)\times \mathrm{Br}(e_b\rightarrow  e_a\bar{\nu}_{a}\nu_b), \label{brlfvdecay2}
\end{equation}
where $\alpha_e\simeq 1/137$,  $F_{(ba)L,R}= \frac{C_{(ba)L,R}}{m_b} \times \left(\frac{g^2e}{32\pi^2 m_W^2}\right)^{-1} $,  and $C_{(ba)L,R}$ are the one-loop contributions originating from the diagrams shown in Fig.~\ref{fig_eitoejgamma}.   The  well-known experimental values of $\mathrm{Br}(e_b\rightarrow  e_a\bar{\nu}_{a}\nu_b)$  are  $\mathrm{Br}(\tau\rightarrow\mu\bar{\nu}_{\mu}\nu_{\tau}) \simeq 17.41\%$,  $\mathrm{Br}(\tau\rightarrow e\bar{\nu}_{e}\nu_{\tau}) \simeq 17.83\%$, and  $\mathrm{Br}(\mu\rightarrow e\bar{\nu}_{e}\nu_{\mu}) \simeq 100\%$~\cite{Tanabashi:2018oca}.  The analytical forms of  $C_{(ba)L,R}$ are  derived based on previous results~\cite{Lavoura:2003xp,Hue:2017lak}. Accordingly, we can use the limit $m_a^2,m_b^2\simeq0$, where the results are  as  follows, 
\begin{align}
\label{eq_FLR}
F_{(ba)L,R} &=F^{(1)}_{(ba)L,R} +F^{(2)}_{(ba)L,R},\crn 
F^{(1)}_{(32)L}&=
   \sum_{i=1}^3\frac{2 m^2_WY^{\sigma^0_1*}_{1i}Y^{\sigma^0_1}_{2i}}{g^2m_{\sigma^0_1}^2} g_s(t_{\sigma^0_1,i})  +\sum_{i=1}^3\frac{2 m_{\mu}m^2_WY^{h_6*}_{1i}Y^{h_6}_{2i}}{m_{\tau}g^2m_{h_6}^2} g_s(t_{h_6,i}),\crn 
 F^{(1)}_{(32)R}&=
   \sum_{i=1}^3\frac{2 m_{\mu}m^2_WY^{\sigma^0_1*}_{1i}Y^{\sigma^0_1}_{2i}}{m_{\tau}g^2m_{\sigma^0_1}^2} g_s(t_{\sigma^0_1,i})  +\sum_{i=1}^3\frac{2 m^2_WY^{h_6*}_{1i}Y^{h_6}_{2i}}{g^2m_{h_6}^2} g_s(t_{h_6,i}),\crn
F^{(1)}_{(b1)L}&= \frac{m_e}{m_b}F^{(1)}_{(b1)R}= 
\sum_{i=1}^3\frac{2 m_{e}m^2_WY^{h_6*}_{3i}Y^{h_6}_{(b-1)i}}{m_{b} g^2m_{h_6}^2} g_s(t_{h_6,i}) , \crn
F^{(2)}_{(32)L}&= \frac{m_\mu}{m_\tau}F^{(2)}_{(32)R} =\frac{2m^2_W}{m^2_{V^0}}\sum_{i=1}^3V'^{E}_{1i}V'^{E*}_{2i} g_v(t_{v,i}), \crn
F^{(2)}_{(b1)L}&= \frac{m_e}{m_b}F^{(2)}_{(b1)R} =\frac{2m^2_W}{m^2_{V^0}}\sum_{i=1}^3V'^{E}_{3i}V'^{E*}_{(b-1)i} g_v(t_{v,i}),
\end{align}
where $t_{x,i}=m^2_{E_i}/m^2_x$  ($x=\sigma^0_1,h_6,V^0$),  
\begin{align}
V'^E_{ai}= \left\{\begin{array}{cc}
(V^E_L)_{ai},& a=3 \\ 
\frac{1}{\sqrt{2}}(V^E_L)_{ai}, & a=1,2 
\end{array} \right.,
\end{align}
and the functions $g_s(t_{s,i})$, $g_v(t_{v,i})$  are derived in  Appendix~\ref{app_CLR}. 

We note that $\sigma^0_1$ only contributes to  LFV decays $t\rightarrow\mu \gamma$ and $h\rightarrow \mu\tau$.  Because of the $\sigma^0_1$ couplings with only $\mu$ and $\tau$. This is the proper property of the flipped 3-3-1 model, where left-handed electron is a component of a sextet, while the $\tau$ and $\mu$ are arranged in triplets as other usual 3-3-1 models.  Consequently, the amplitudes of the two decays $h\rightarrow \mu\tau$ and $\tau \rightarrow \mu \gamma$   receive more one-loop contributions than the remaining decay amplitudes, hence we expect that the Br$(h\rightarrow \tau\mu) $ and Br$(\tau\rightarrow \mu\gamma)$ will be large.   

\subsection{\label{numerical} Numerical discussions}
In this numerical discussion,  the  unknown input parameters are:  the masses and mixing parameters of the heavy leptons $s^E_{ij}$ and $m_{E_i}$;  heavy neutral Higgs masses and mixing $m_{\sigma^0_1}$, $m_{h_6}$ and $s_{2s}$. In addition,  the unknown VEVs in the model are  $k_1$ and $n_2$. From Eqs.~\eqref{eq_h06} and \eqref{eq_mgauge},  we have  
\begin{align}\label{eq_nS}
n_S&= \frac{s_{2s}n_2}{c_{2s}\sqrt{2}}, \quad n_2^2(1+2t^2_{2s})=\frac{(3-4s_W^2) m^2_{Z'}}{4g^2c_W^2},
\end{align}
where $t_{2s}\equiv s_{2s}/c_{2s}$.  
This means that $n_2^2 +4 n_S^2\simeq (2.15 m_{Z'})^2$.  For the latest lower bound of $m^2_{Z'}\ge 4 $ TeV reported from experiment~\cite{Aaboud:2017sjh}, we have $ \sqrt{n_2^2 +4 n_S^2}\ge 8.3$ TeV.  For our numerical investigation in this work, we will fix $ \sqrt{n_2^2 +4 n_S^2}= 8.3$ TeV, $n_2=1$ TeV, $n_S\geq 4$ TeV, leading to $t_{2s}=\sqrt{2}n_S/n_2=4\sqrt{2}$; equivalently $s_{2s}\simeq 0.985$.  The large $s_{2s}$ corresponds to the large  Yukawa coupling $Y^{h_6}$ given in Eq.~\eqref{eq_Ys}. Because $k_1$ generates masses for the lepton $\tau$ at the tree level, it should not be too small. In addition,  $\mu^2_{12}$ given in Eq.~\eqref{eq_minEqHapp} is too large  if $k_1$ is too small. Hence we will choose that $10\;\mathrm{GeV}\leq k_1\leq50$ GeV.  The above particular choice of $m_{E_i}$  is an illustration for a general consideration where large Br$(h\rightarrow e_be_a)$ needs $m_{E_i}-m_{E_j}= \mathcal{O}(10^2)$ GeV  when $m_{E_i}=\mathcal{O}(1)$ TeV is applied in our discussion. 

In the first numerical investigation, the default values of the  inputs are $k_1=20$ GeV, $\lambda_{13}=1$, $f^{\phi}=2$ TeV, $m_{E_1}= 1$ TeV,  $m_{E_k}=m_{E_1} -k \times100$ GeV, $n_2=1$ TeV, $s_{2s}=0.985$, $m_{\sigma^0_1}=m_{h_6}=1$ TeV.   The perturbative limit of the Yukawa couplings relating to heavy lepton masses  gives $m_{E_1}\leq n_2 \sqrt{4\pi}=3.5$ TeV for $n_2=1$ TeV. Values of $m_{E_{2,3}}$ are chosen to avoid the degenerate masses of the three charged heavy leptons which result in   Br$(e_b\rightarrow e_a\gamma)=0$.   

All other well-known parameters are taken from Ref.~\cite{Tanabashi:2018oca}, namely  the Higg boson mass and its total decay width $m_h=125.01$ GeV and  $\Gamma_h=4.07\times 10^{-3}$ GeV; the  mass of the $W$  boson,   the masses of normal leptons $m_e,\; m_{\mu}, \; m_{\tau}$, the gauge couplings and $\alpha_e$. 

\begin{figure}[ht]
	\centering
	\begin{tabular}{cc}
		\includegraphics[width=7.5cm]{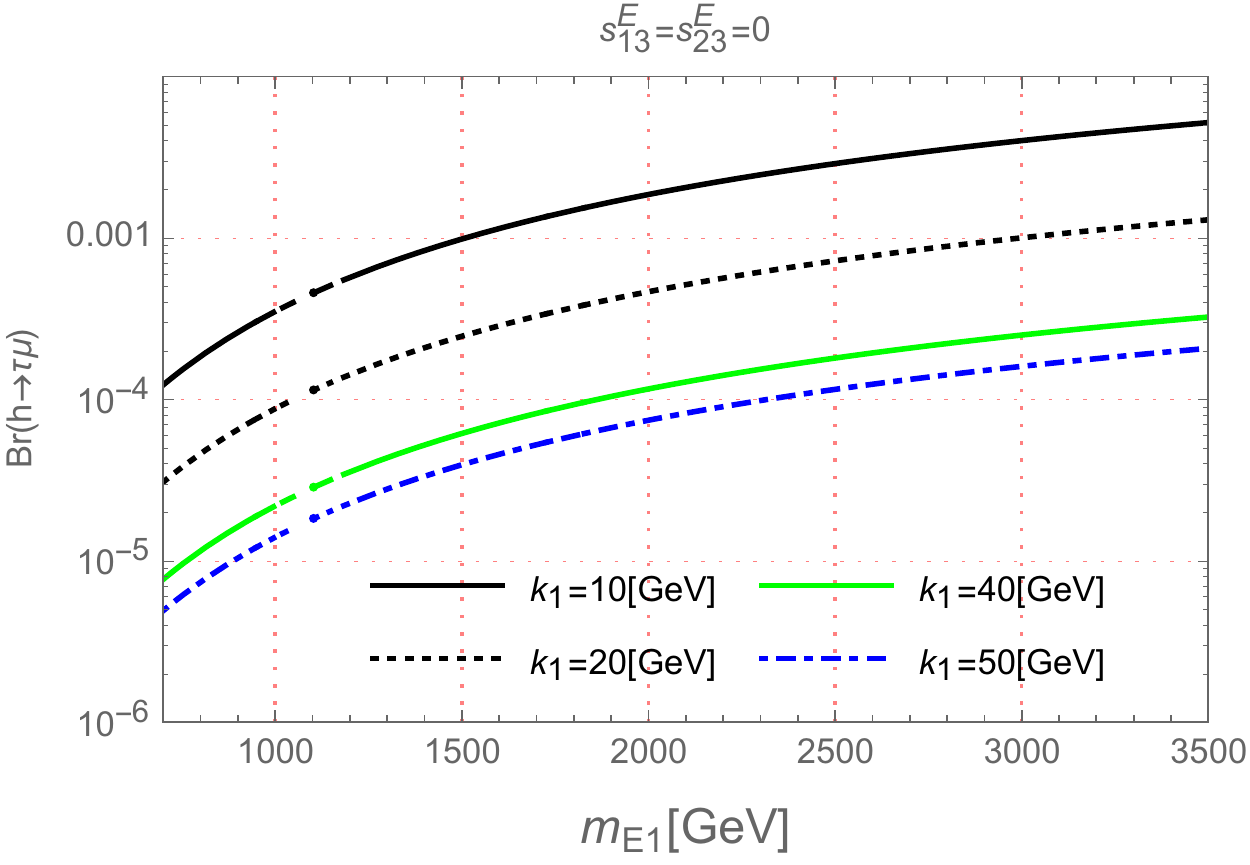}& \includegraphics[width=7.5cm]{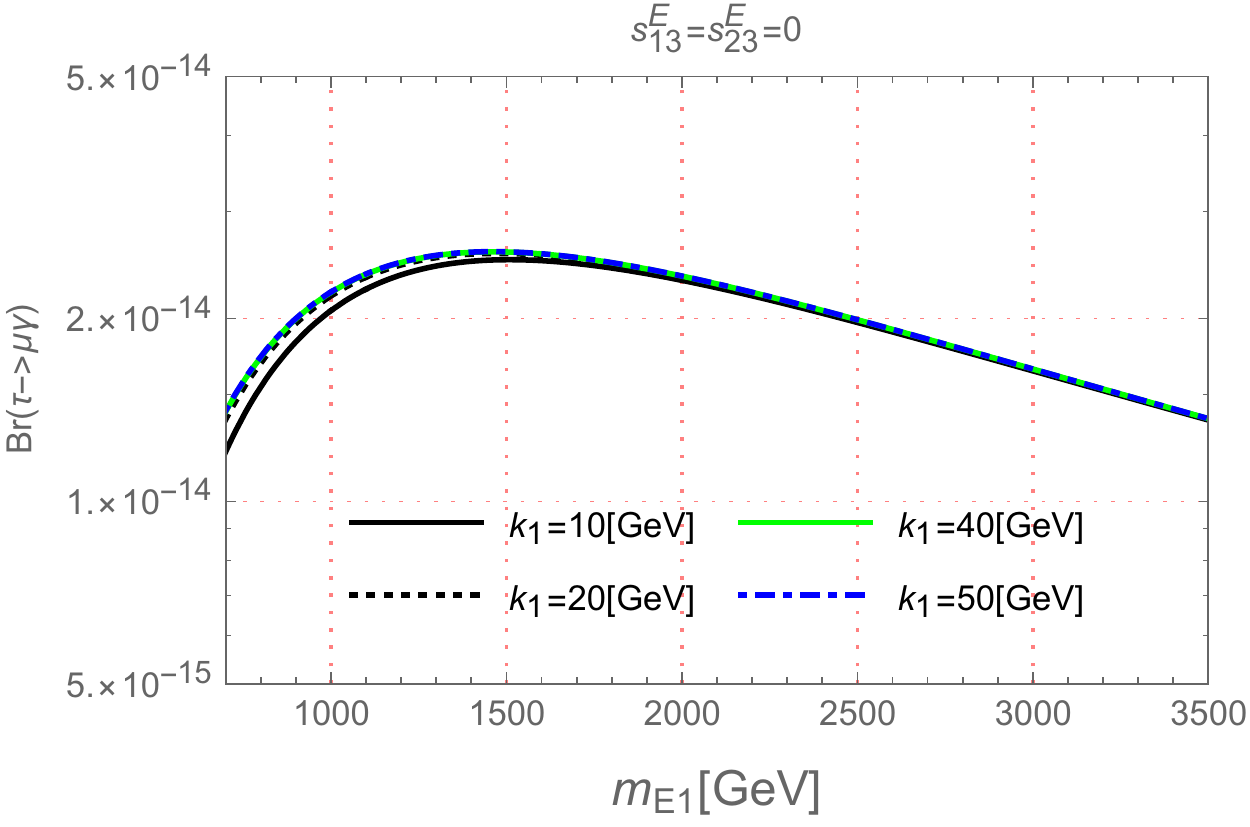}
		\\
	\end{tabular}
	\caption{$\mathrm{Br}(h\rightarrow \tau \mu )$ and $\mathrm{Br}(\tau\rightarrow \mu\gamma)$ as functions of $m_{E_1}$ in the cases $s^E_{12}=\frac{1}{\sqrt{2}}$ and $s^E_{13}=s^E_{23}=0$.}\label{fig_s12mt}
\end{figure}

Regarding the mixing matrix $V^E_L$,  we first consider three cases of only one of  $s^E_{ij}=1/\sqrt{2}$, which correspond to the maximal mixing of only two heavy charged leptons.  Hence, these result in large branching ratios of  some of the LFV decays while the remaining ones vanish. This help us to estimate the largest  branching ratios  of LFV decays.  In the case of $s_{12}=1/\sqrt{2}$ and  $s_{13}=s_{23}=0$, we always have $\mathrm{Br}(h\rightarrow \mu e)=\mathrm{Br}(h\rightarrow \tau e)=\mathrm{Br}(\mu\rightarrow e\gamma)=\mathrm{Br}(\tau\rightarrow e\gamma)=0$.  In contrast, the $\mathrm{Br}(h\rightarrow \tau \mu )$ and $\mathrm{Br}(\tau\rightarrow \mu\gamma)$ as functions of $m_{E_1}$ with different fixed $k_1$ are shown in Fig.~\ref{fig_s12mt}. 
It can be seen that Br$(\tau\rightarrow \mu\gamma)$ is much smaller than the current experimental bound given in Eq.~\eqref{eq_ebagaex}.  Although the Br$(h\rightarrow \tau\mu)\sim \mathcal{O}(10^{-3})$ is close to the current experimental bound in Eq.~\eqref{eq_ebagaex}, the lower bounds obtained from near-future experiments can be used to constrain the parameter space. The two parameters $k_1$ and $m_{E_1}$ strongly affect  on Br$(h\rightarrow \tau\mu)$ but Br$(\tau\rightarrow \mu\gamma)$ depends weakly on them. This property can be explained as follows. The dominant contribution to the $h\rightarrow \tau\mu$ decay amplitude is $\Delta^{h_6\sigma^0_1}_{32}$, which is  proportional to $f^{\phi}m_{\tau} m^2_{E_i}/ k_1$ and $C_0\sim 1/m^2_{E_i}$ for $m^2_{E_i}\gg m^2{h_6},m^2_{\sigma^0_1}$. For the decay amplitude $\tau\rightarrow \mu\gamma$ the contribution relating to $\sigma^0_1$ is much smaller than that relating to $h_6$. 

\begin{figure}[ht]
	\centering
	\begin{tabular}{cc}
		\includegraphics[width=7.5cm]{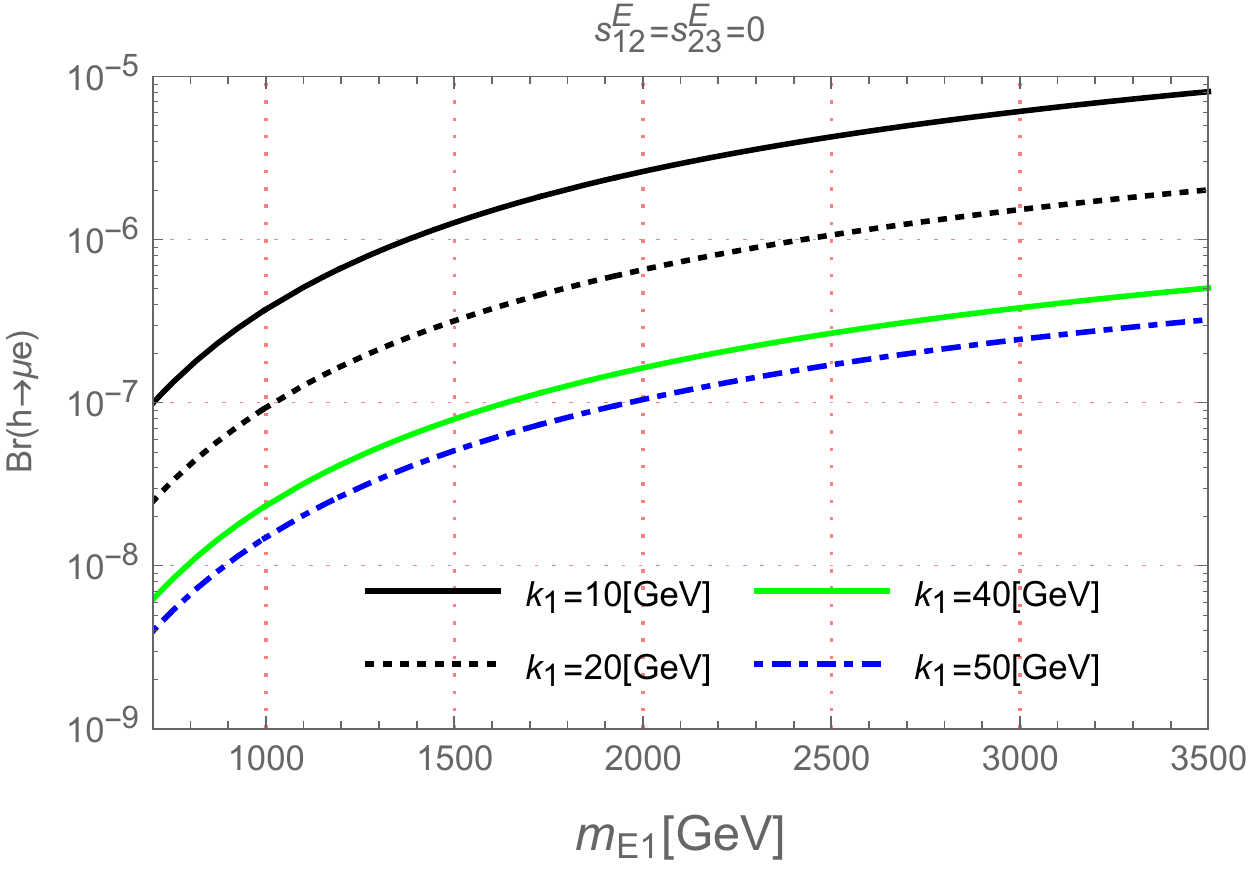}& \includegraphics[width=7.5cm]{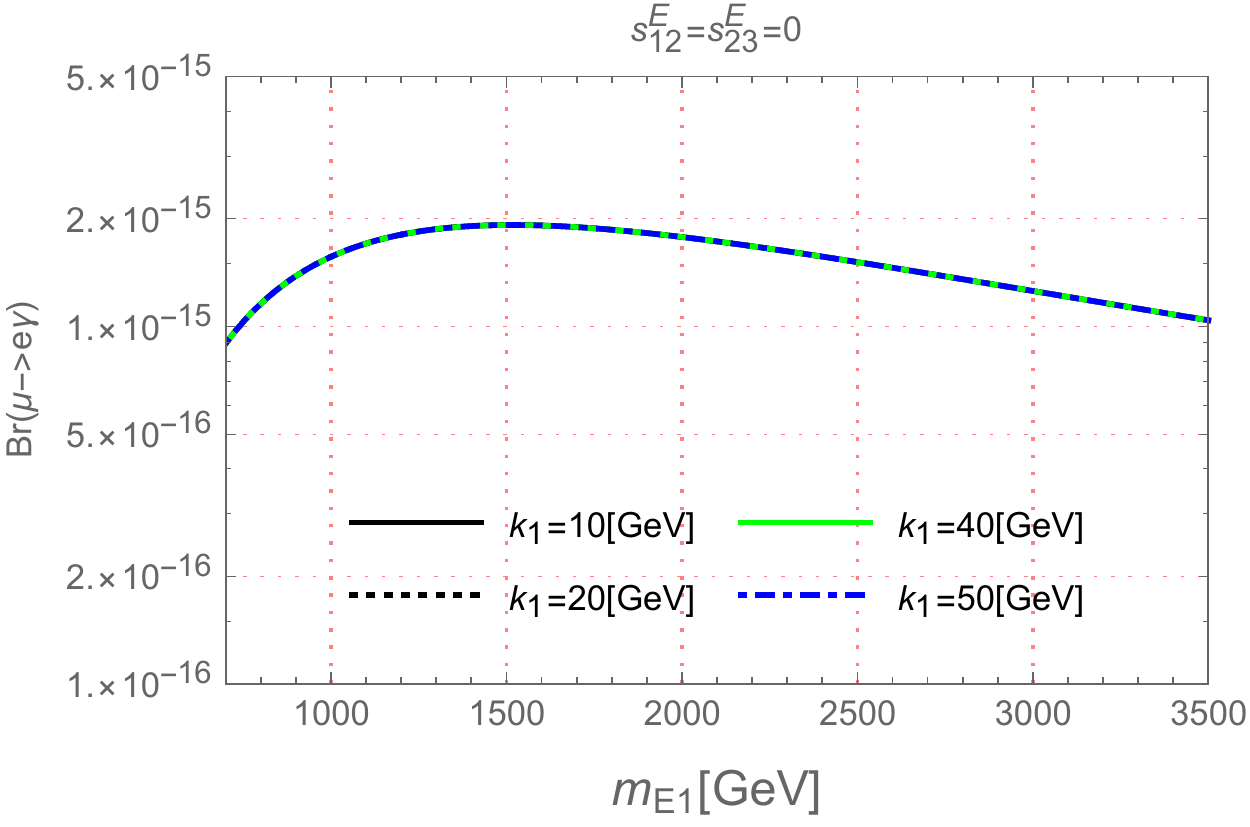}
		\\
	\end{tabular}
	\caption{$\mathrm{Br}(h\rightarrow \mu e)$ and $\mathrm{Br}(\mu\rightarrow \,e\gamma)$ as functions of $m_{E_1}$ in the case $s^E_{13}=\frac{1}{\sqrt{2}}$ and $s^E_{12}=s^E_{13}=0$.}\label{fig_s13me}
\end{figure}

Similarly, with $s^E_{12}=s^E_{23}=0$ and $s^E_{13}=\frac{1}{\sqrt{2}}$,  we have only two non-zero  $\mathrm{Br}(h\rightarrow  \mu e)$ and $\mathrm{Br}(\mu\rightarrow e\gamma)$. Illustrations of these branching ratios as functions of $m_{E_1}$ with different fixed $k_1$ are shown in Fig.~\ref{fig_s13me}.
Accordingly, Br$(\mu\rightarrow e\gamma)\leq \mathcal{O}(10^{-15})$, which  still satisfies the lower bound in Eq.~\eqref{eq_ebagaex}. It is noted that although Br$(h\rightarrow \mu e)$ is sensitive to $k_1$, the Br$(\mu\rightarrow e\gamma)$ is not, because it does not receive contribution from Yukawa coupling of $\sigma^0_1$. 

\begin{figure}[ht]
	\centering
	\begin{tabular}{cc}
		\includegraphics[width=7.5cm]{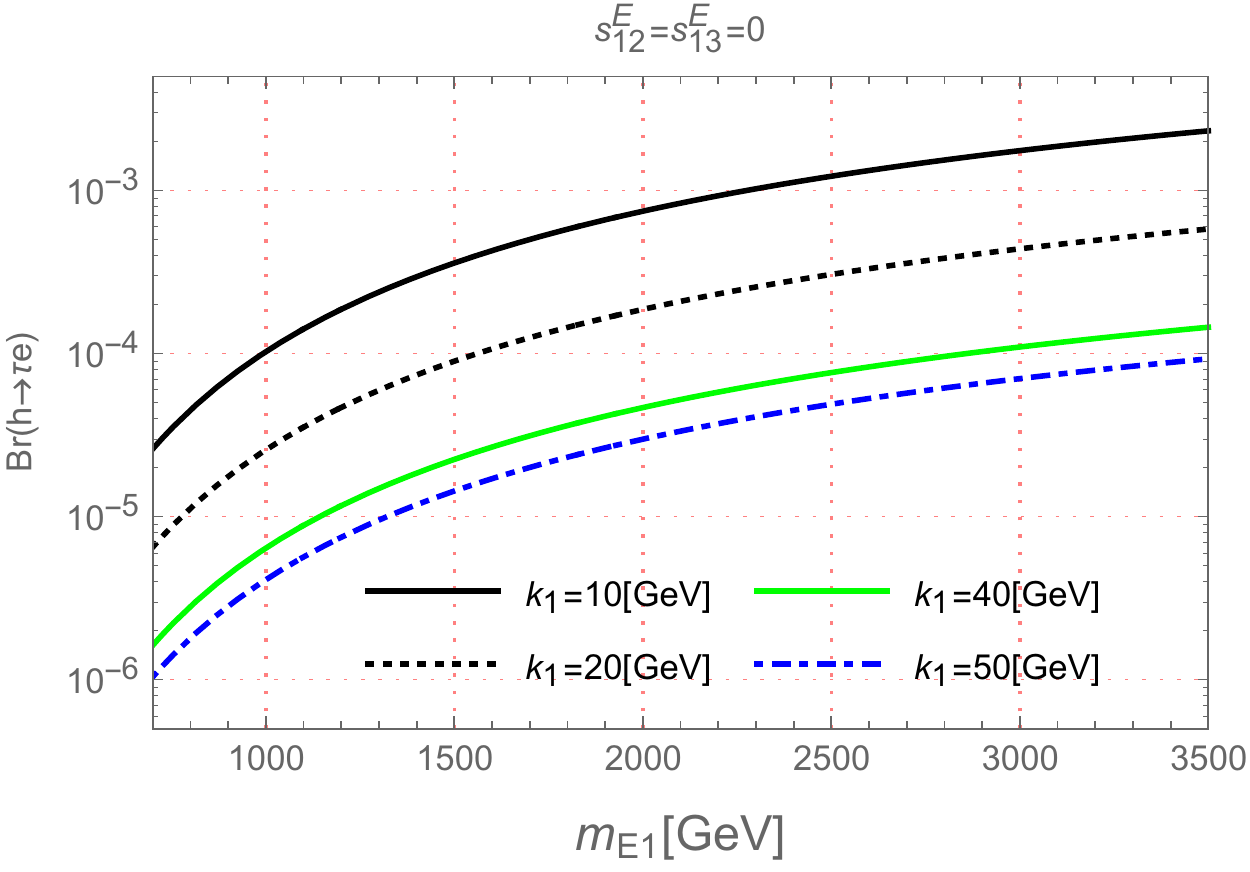}& \includegraphics[width=7.5cm]{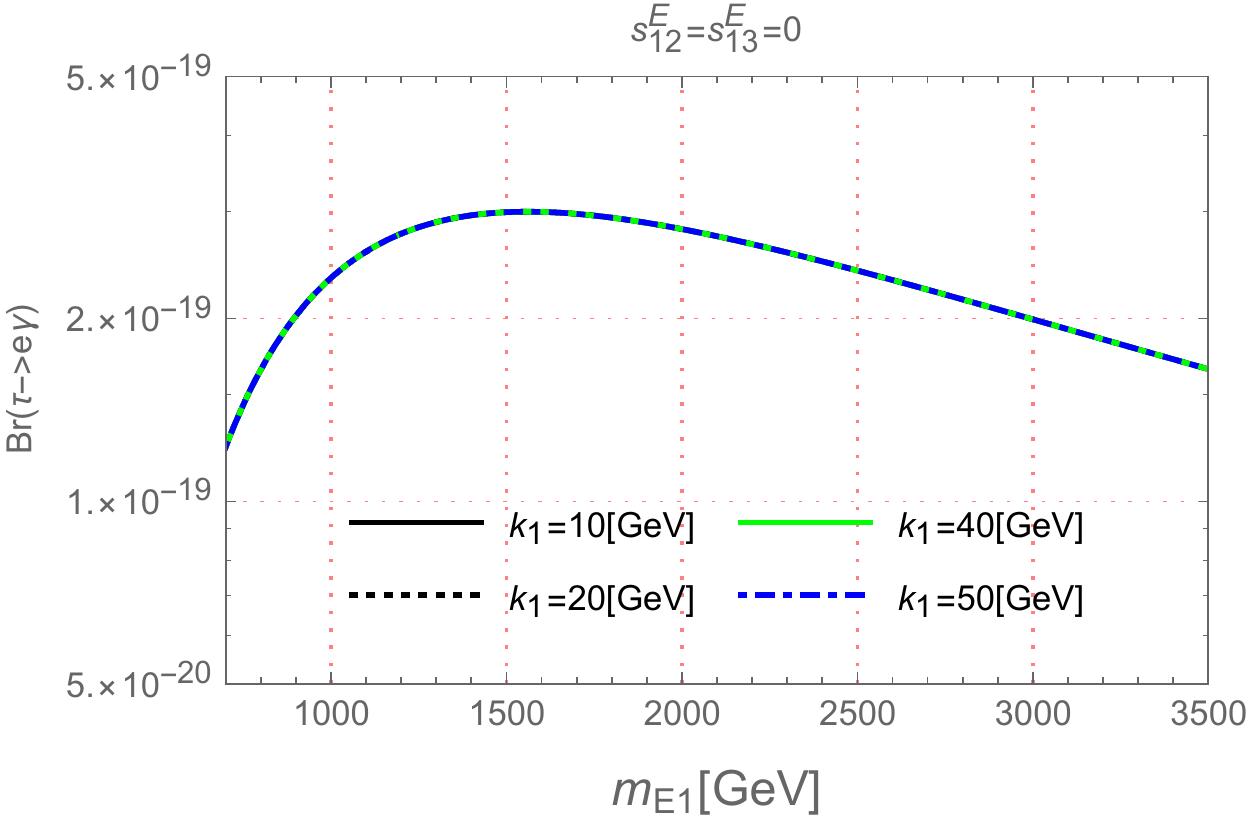}
		\\
	\end{tabular}
	\caption{$\mathrm{Br}(h\rightarrow \tau e)$ and $\mathrm{Br}(\tau\rightarrow \,e\gamma)$ as functions of $m_{E_1}$ in the case $s^E_{23}=\frac{1}{\sqrt{2}}$ and $s^E_{12}=s^E_{13}=0$.}\label{fig_s23te}
\end{figure}

The case of $s^E_{12}=s^E_{13}=0$ and $s^E_{23}=\frac{1}{\sqrt{2}}$ correspond to the  two non-zero  $\mathrm{Br}(h\rightarrow  \tau e)$ and $\mathrm{Br}(\tau\rightarrow e\gamma)$. Illustrations of these branching ratios as functions of $m_{E_1}$ with different fixed $k_1$ are shown in Fig.~\ref{fig_s23te}.
 In this case, Br$(h\rightarrow \tau e)$ has the same order as Br$(h\rightarrow \tau\mu)$ because both of them get dominant contributions from $\Delta^{\sigma^0_1h_6}_{(ba)R}$.  Other contributions to $\Delta_{(ba)}$ have been checked numerically and shown to be very suppressed.  Similarly, the case of Br$(\tau \rightarrow \mu\gamma)$, Br$(\tau \rightarrow e\gamma)$ is much smaller than the current and upcoming experimental sensitivities.

\begin{figure}[ht]
	\centering
	\begin{tabular}{cc}
		\includegraphics[width=7.5cm]{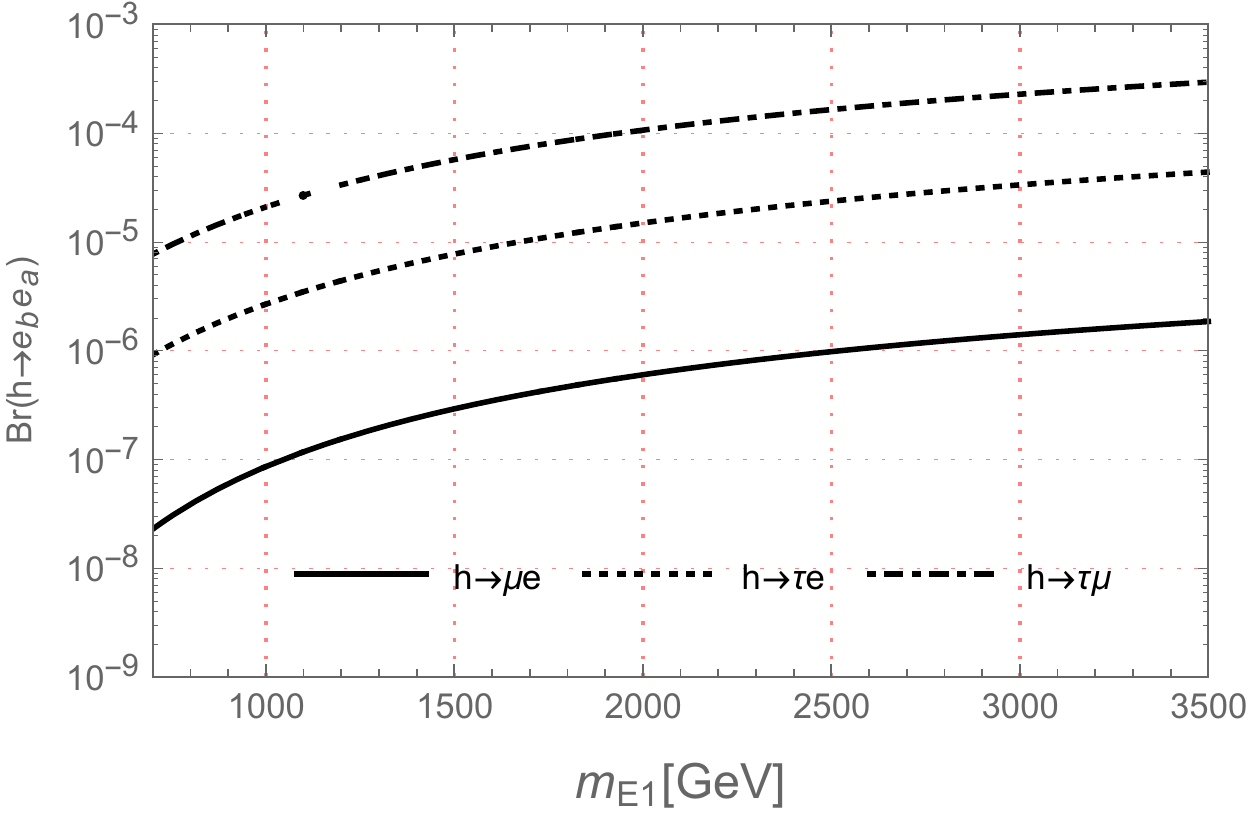}& \includegraphics[width=7.5cm]{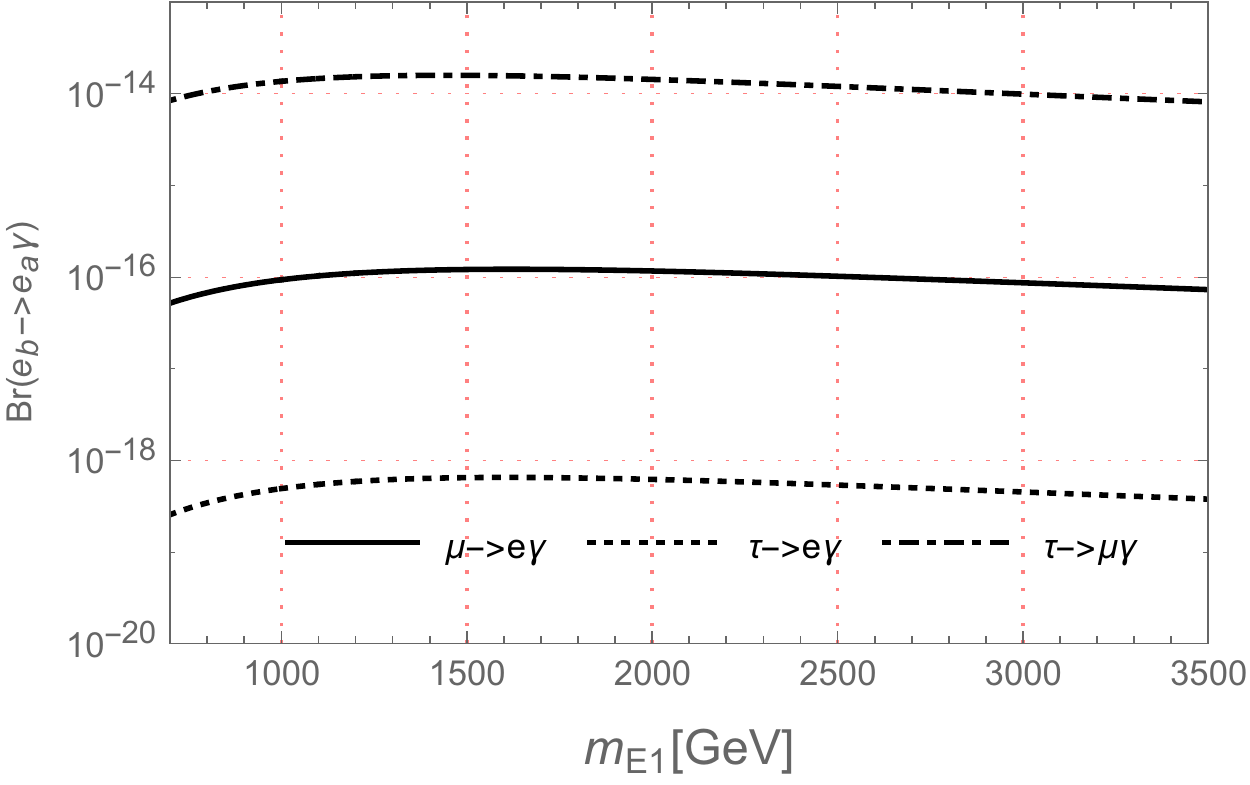}
		\\
	\end{tabular}
	\caption{$\mathrm{Br}(h\rightarrow e_be_a)$ and $\mathrm{Br}(e_b\rightarrow \,e_a\gamma)$ as functions of $m_{E_1}$ in the case $s^E_{ij}=\frac{1}{\sqrt{2}}$ for all $i,j=1,2,3$ and $i<j$.}\label{fig_LFVmE1}
\end{figure}

In order to illustrate the effects of heavy lepton masses on the magnitude of different LFV decays, we consider the case of  all equal non-zero $s^E_{ij}=\frac{1}{\sqrt{2}}$.  The branching ratios of all LFV decays are functions of $m_{E_1}$,  numerical illustrations of which are shown in Fig.~\ref{fig_LFVmE1}.

\begin{figure}[ht]
	\centering
	\begin{tabular}{cc}
		\includegraphics[width=7.5cm]{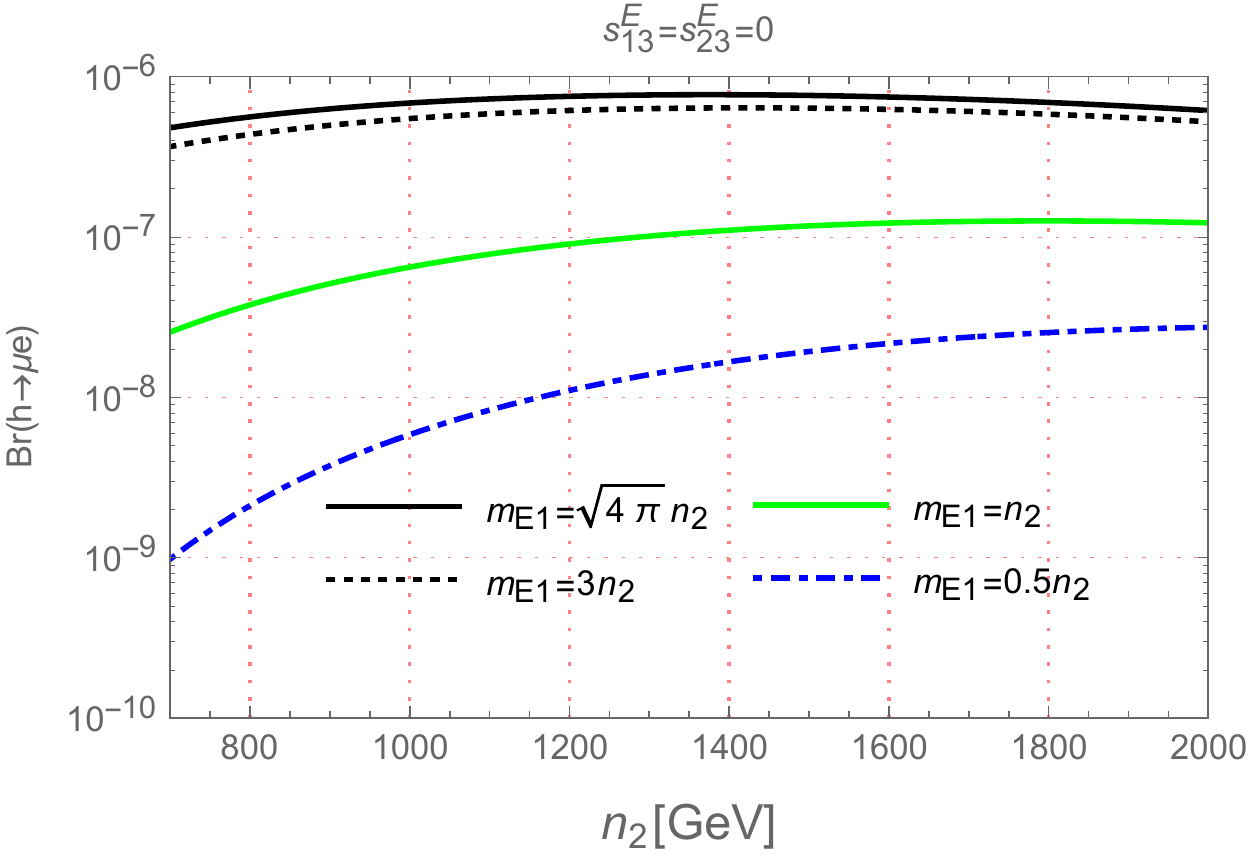}& \includegraphics[width=7.5cm]{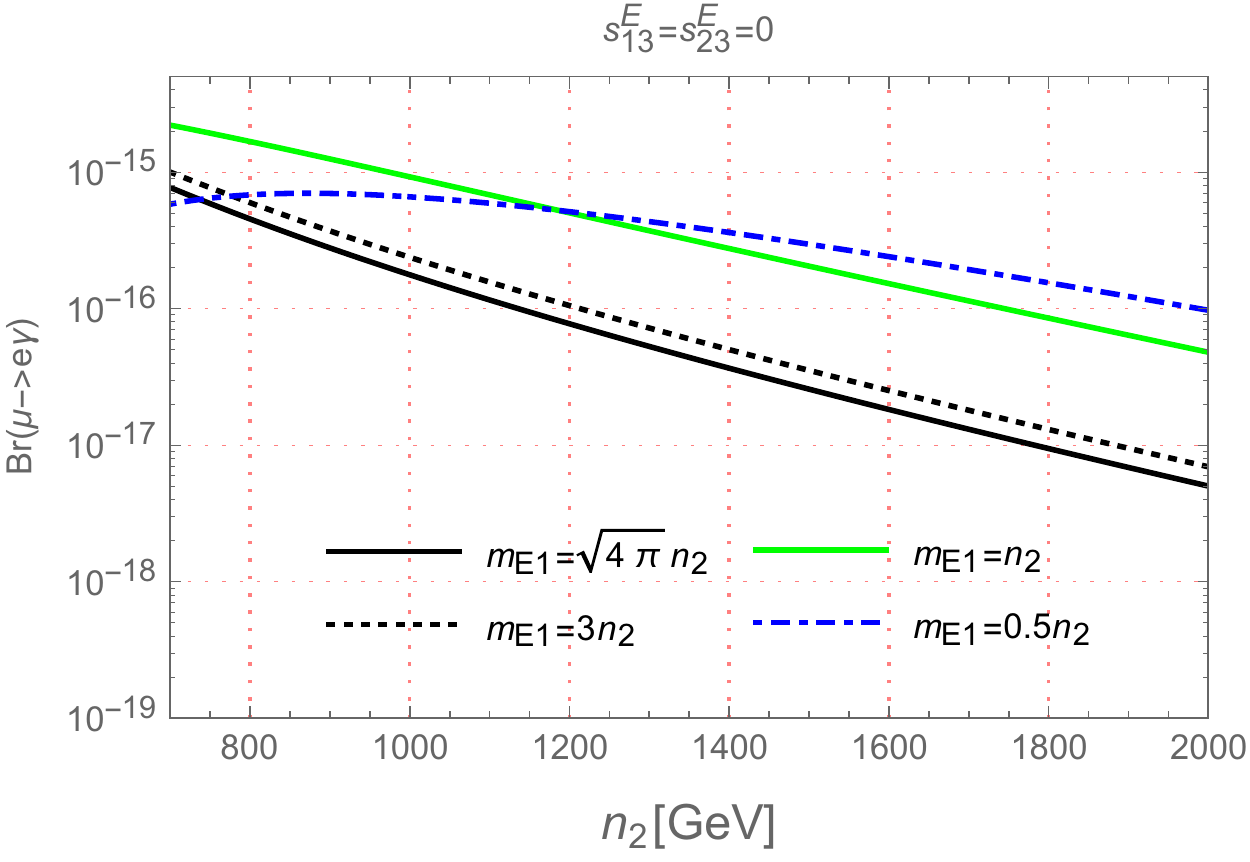}
		\\
	\end{tabular}
	\caption{$\mathrm{Br}(h\rightarrow \mu e)$ and $\mathrm{Br}(\mu\rightarrow \,e\gamma)$ as functions of $n_{2}$ in the case of  $m_{Z'}=4$ TeV, $s^E_{13}=\frac{1}{\sqrt{2}}$ and $s^E_{12}=s^E_{23}=0$.}\label{fig_Brmen2}
\end{figure}

We consider a region with large Br$(\mu\rightarrow e\gamma)$, where the necessary conditions are large $s^E_{13}$, small values of $n_2$ and small $m_{h_6}$.  The illustration is shown in Fig.~\ref{fig_Brmen2}, where we fix $m_{Z'}=4$ TeV and $m_{h_6}=500$ GeV, then plot branching ratios of  LFV decays as functions of $n_2$ with different $m_{E_1}/n_2\leq \sqrt{4\pi}$ satisfying the perurbative limit.  
We can see again that Br$(\mu\rightarrow e\gamma) \leq \mathcal{O}(10^{-15})$. The large Br$(\mu \rightarrow e\gamma)$ corresponds to the regions of small $n_2$ and small $m_{h_6}$. 

\begin{figure}[ht]
	\centering
	\begin{tabular}{cc}
		\includegraphics[width=7.5cm]{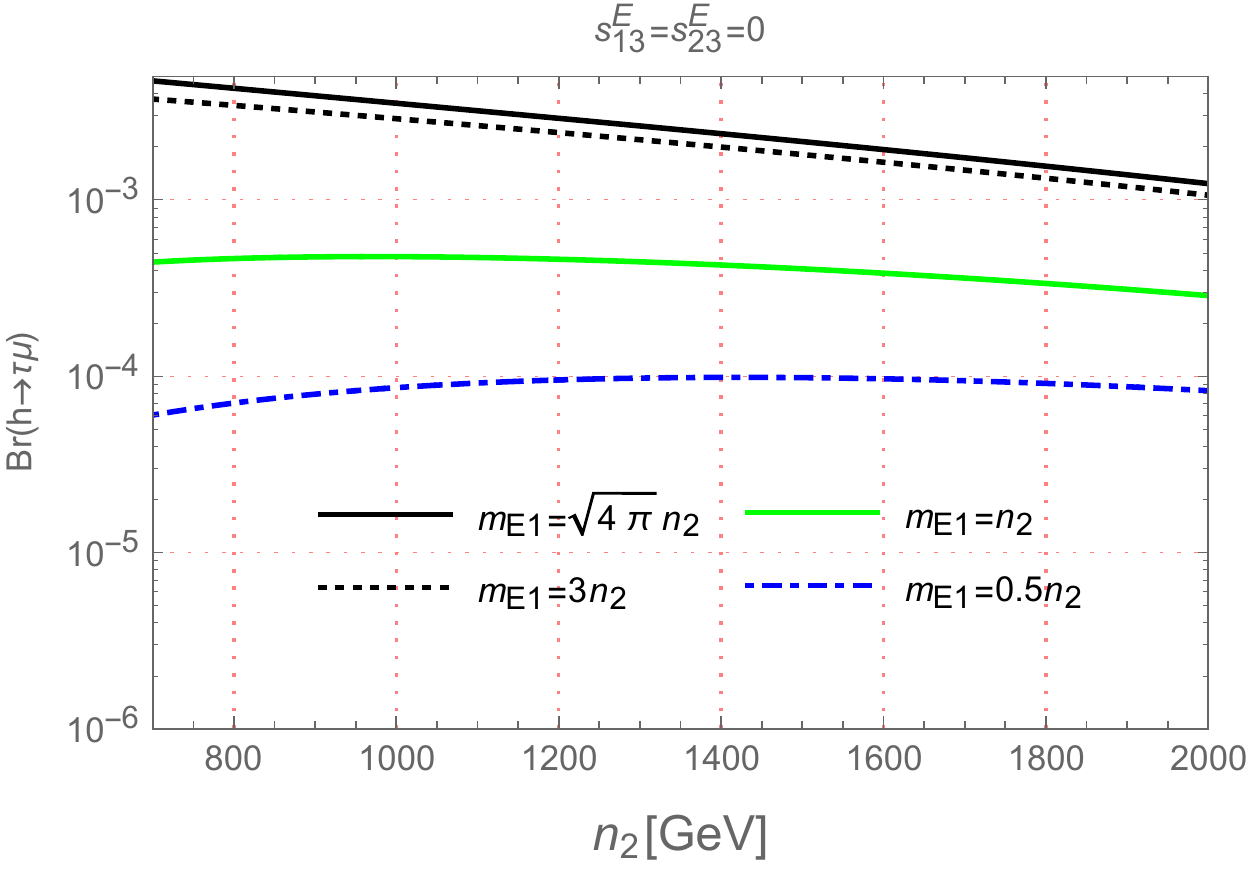}& \includegraphics[width=7.5cm]{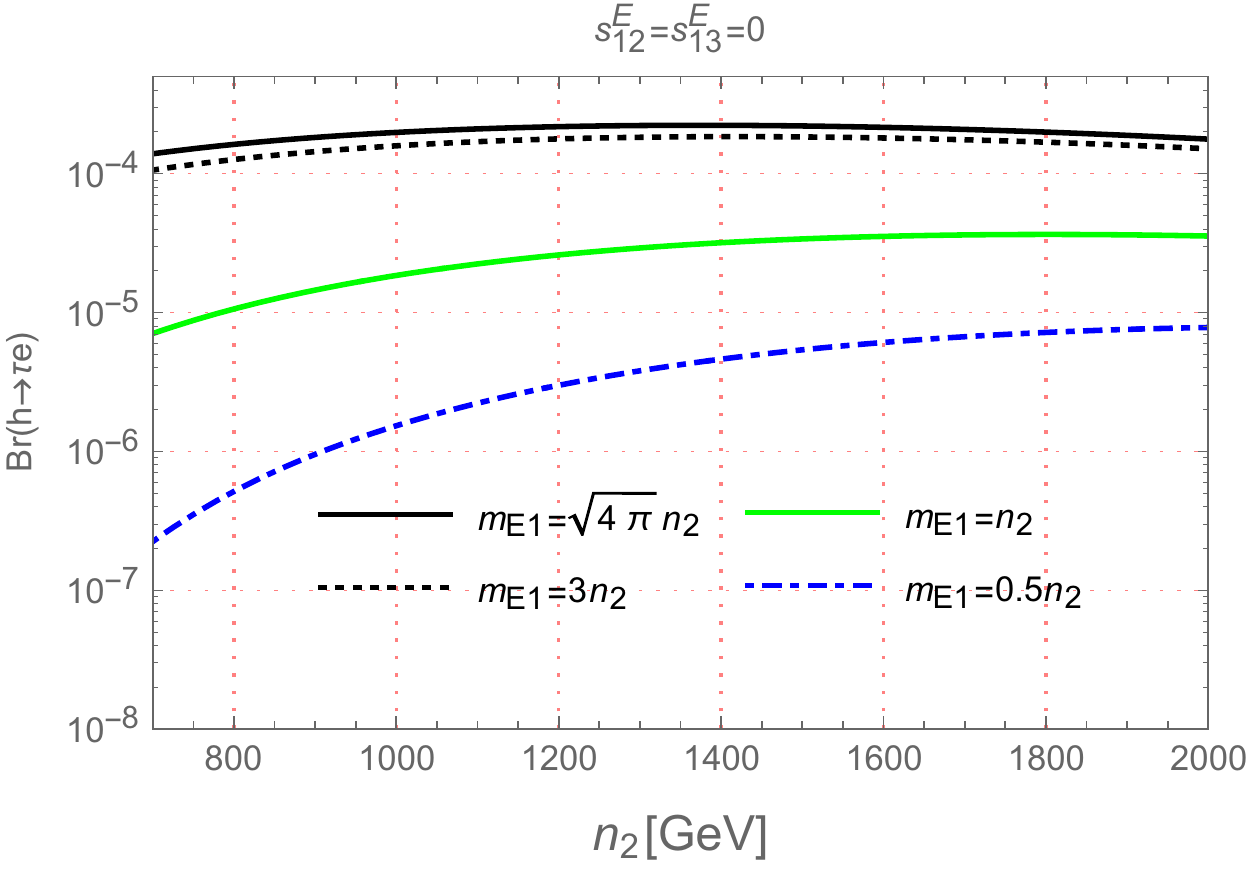}
		\\
	\end{tabular}
	\caption{$\mathrm{Br}(h\rightarrow\tau \mu)$ (left) and $\mathrm{Br}(h\rightarrow\tau e)$ (right) as functions of $n_{2}$ in the case of $m_{Z'}=4$ TeV and  $s^E_{12}=\frac{1}{\sqrt{2}}$ ($s^E_{23}=\frac{1}{\sqrt{2}}$).}\label{fig_Brheab}
\end{figure}

Similarly, the $\mathrm{Br}(h\rightarrow\tau \mu)$ and $\mathrm{Br}(h\rightarrow\tau e)$ as functions of $n_2$ are shown in Fig.~\ref{fig_Brheab}. The Br$(\tau\rightarrow \mu\gamma, e\gamma)$ are much smaller than current experimental constraints so we do not show them again. 
We just mention here a property  that all Br$(h\rightarrow e_be_a)$ are enhanced with increasing $m_{E_1}$, which has an upper bound originating from the perturbative limit of the Yukawa couplings.  Hence  the upper bounds of Br$(h\rightarrow e_be_a)$ correspond to the largest values of the Yukawa couplings. In contrast, all Br$(e_b \rightarrow e_a \gamma)$ decrease with increasing $m_{E_1}$ when $n_2$ is large enough.  

\begin{figure}[ht]
	\centering
	\begin{tabular}{cc}
		\includegraphics[width=7.5cm]{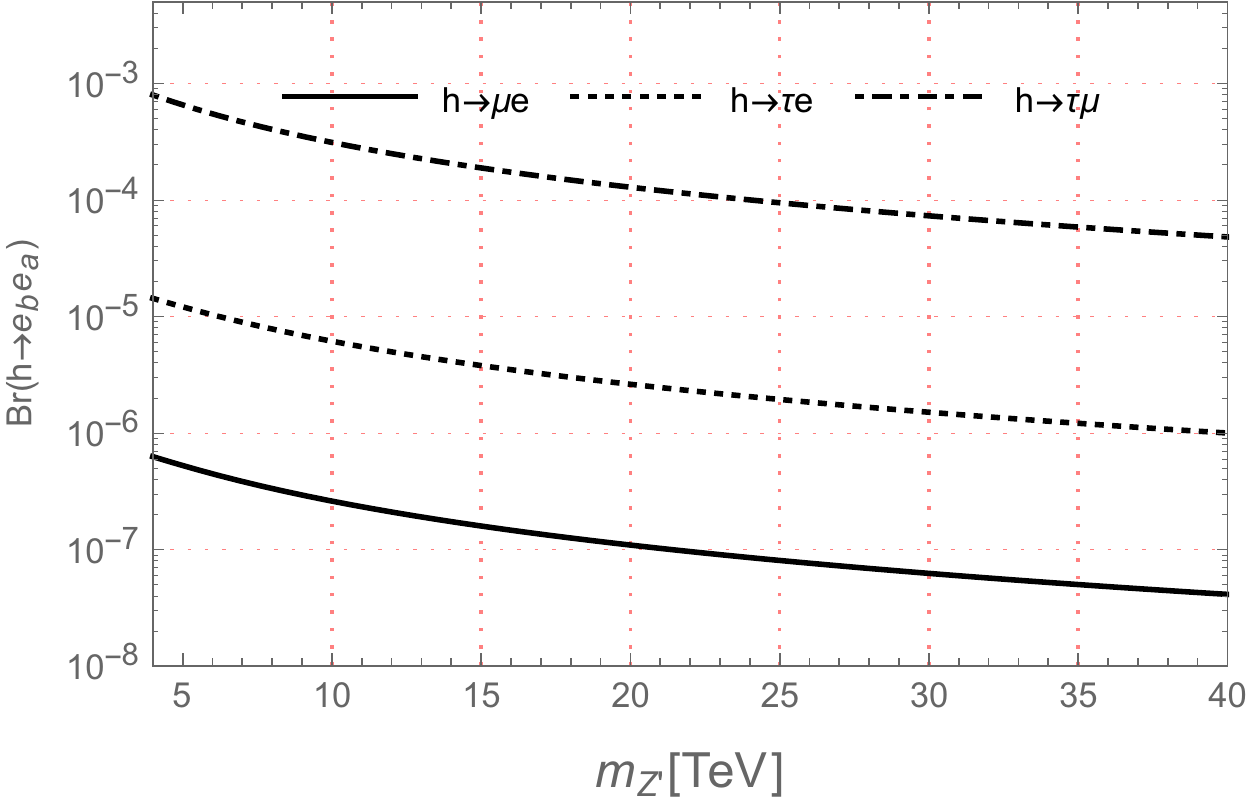}& \includegraphics[width=7.5cm]{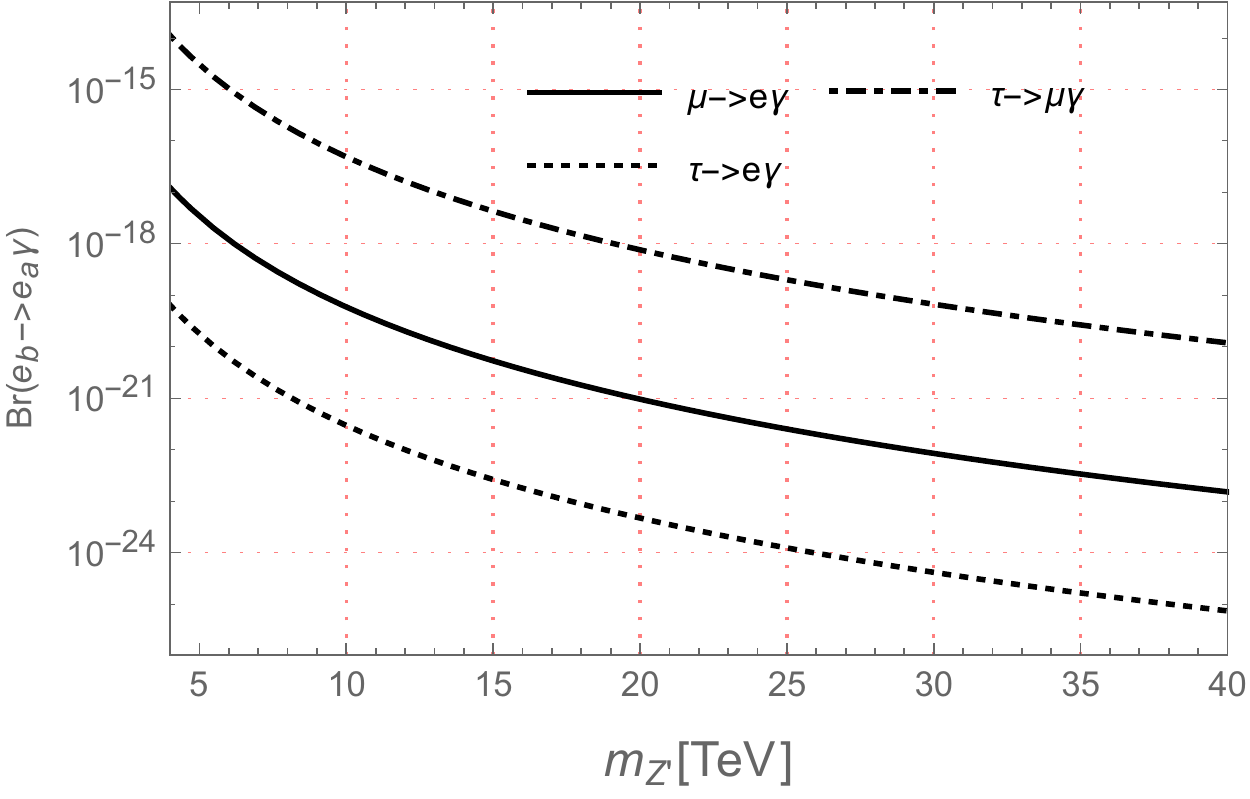}
		\\
	\end{tabular}
	\caption{$\mathrm{Br}(h\rightarrow e_be_a)$ (left) and $\mathrm{Br}(e_b\rightarrow e_a\gamma)$ (right) as functions of $m_{Z'}$ in the case of $n_2=m_{Z'}/4$ and   $s^E_{ij}=\frac{1}{\sqrt{2}}$ with all $i<j$, $i,j=1,2,3$.}\label{fig_LFVmZp}
\end{figure}

To estimate how large the LFV branching ratios can beome when  $m_{Z'}$ is large, we  fix $n_2=m_{Z'}/4 \geq 1$ TeV, then $t_{2s}$ and $n_S$ are determined from the relations given in Eq.~\eqref{eq_nS}.  The Br of LFV decays as functions of $m_{Z'}$ are illustrated in Fig.~\ref{fig_LFVmZp}.
In this case  we can see that all LFV branching ratios  decrease with larger $m_{Z'}$, but  Br$(h\rightarrow \tau\mu)$ and Br$(h\rightarrow \tau e)$ are still close to the order of $O(10^{-5})$ or larger. Hence these decay channels are still interesting for experiments. On the other hand, all Br$(e_b\rightarrow e_a\gamma)$ decrease rapidly with increasing $m_{Z'}$. They will not be detected by upcoming experiments. 

\begin{figure}[ht]
	\centering
	\begin{tabular}{cc}
		\includegraphics[width=4cm]{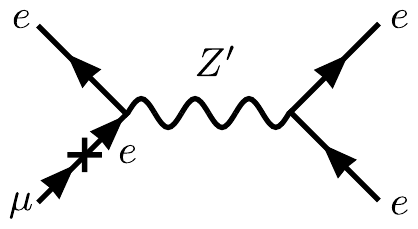} & \quad\quad\quad  \includegraphics[width=11cm]{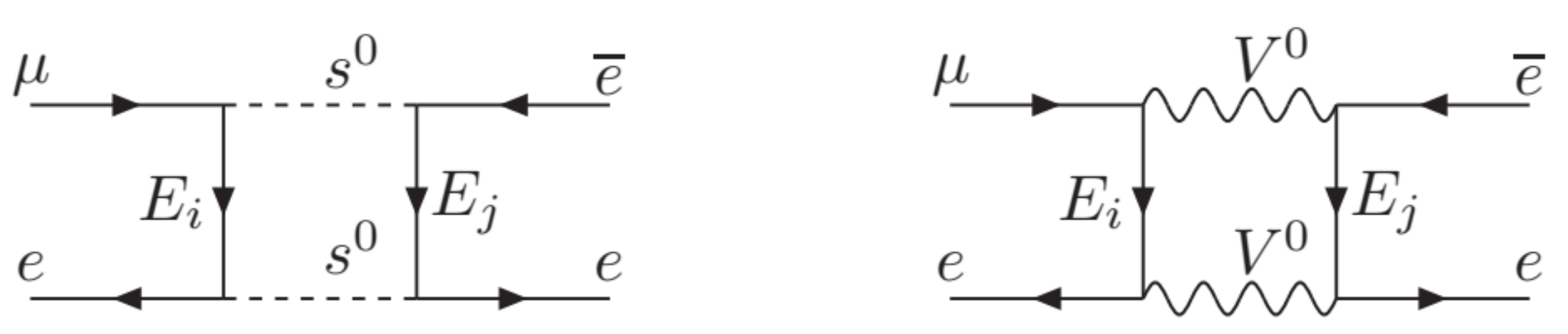}
		\\
	\end{tabular}
	\caption{Tree level~\cite{Fonseca:2016tbn} and one loop Feynman diagrams contributing to the decay $\mu\rightarrow e\overline{e}e$.}\label{fig_muto3e}
\end{figure}

Apart from the  LFV decay Br$(\mu\rightarrow e\gamma)$,  the  LFV decay $\mu \rightarrow e\overline{e}e$ is also highly constrained from experimental data, Br$(\mu \rightarrow e\overline{e}e)<\mathcal{O}(10^{-12})$~\cite{Bellgardt:1987du}. A discussion in ref.~\cite{Fonseca:2016tbn} showed that there exists a tree-level contribution from  the heavy gauge boson $Z'$ to this decay amplitude; see the first Feynman diagram in Fig.~\ref{fig_muto3e}. 
Accordingly, the experimental upper bound of  $(\mu \rightarrow e\overline{e}e)<\mathcal{O}(10^{-12})$ was shown to give a constraint of $m_{Z'}\ge 3$ TeV, which is less strict than that obtained from LHC. In addition, there appear  one-loop contributions to this decay because of the same LFV couplings as those  result in the LFV decay $\mu\rightarrow e\gamma$, see the second and third diagrams in Fig.~\ref{fig_muto3e}. From previous works~\cite{Abada:2014kba, Lindner:2016bgg},  it can be seen  that the one-loop contributions to the two mentioned LFV decays are of same orders. Therefore, the  numerical investigations on the Br$(\mu\rightarrow e\gamma)$ show that the tree-level contribution of $Z'$ to $\mu\rightarrow e\overline{e}e$ is still dominant, and can be used to constrain the $m_{Z'}$. 
 
\section{Conclusions}
\label{sec_conclusion}
We have investigated LFV decays of the SM-like Higgs boson $h\rightarrow e_be_a$ and charged leptons $e_b \rightarrow e_a\gamma$ in the framework of the flipped 3-3-1 model. The Higgs potential was considered in a simple case, where we  have shown that the model contains an SM-like Higgs boson  that can be identified as the one found experimentally.  The main LFV sources originate from the heavy charged leptons. Because electron is arranged in a sextet, which is different from the two other charged leptons $\tau$ and $\mu$, one-loop contributions to the LFV amplitudes of the decays $h\rightarrow \mu\tau$ and $\tau\rightarrow \mu \gamma$ are  larger than the remaining $h\rightarrow \tau e, \mu e$ and $\tau, \mu\rightarrow e\gamma$, respectively.  Assuming that all new heavy particles are in the TeV scale, the Br$(h\rightarrow\tau\mu,\tau e)$ and  Br$(h\rightarrow\mu e)$  can reach the orders of $\mathcal{O}(10^{-3}-10^{-4})$, and $\mathcal{O}(10^{-6})$, respectively.  These values are very close to the recent lower bounds reported by experiments, and they should be considered for constraining the parameter space of the model if improved lower bounds on these decay rates are published. The large  values of Br for LFVHD still appear even with heavy $m_{Z'}\sim \mathcal{O}(10)$ TeV. On the other hand, the Br$(e_b\rightarrow e_a\gamma)$ always satisfies the current experimental constraints. In addition, our numerical investigation shows that Br$(\tau\rightarrow \mu\gamma,e\gamma)\leq \mathcal{O}(10^{-14})$, which is much smaller than the planned sensitivities of upcoming experiments.  Similarly, Br$(\mu\rightarrow e\gamma)$ can reach the order  of $\mathcal{O}(10^{-15})$  which is more promising for  searching by experiments. 

\section*{Acknowledgments}
This reseach is funded by the An Giang University under Grant No. 19.02.TB. 
\appendix
\section{\label{app_CLR} One loop contribution to the decay amplitudes $e_b\rightarrow e_a \gamma$}
The one-loop contributions to the decays $e_b\rightarrow\,e_a \gamma$ is calculated based on the notations of the PV functions defined in ref.~\cite{Hue:2017lak}. 
\begin{align}
\label{eq_CLR}
C^{\sigma^0_1}_{(32)L}&= \sum_{i=1}^3\frac{-m_{\tau}Q_{E}Y^{\sigma^0_1*}_{1i}Y^{\sigma^0_1}_{2i}}{16\pi^2} \left[ C_1([p^2_i]; m^2_{\sigma^0_1}, m^2_{E_i}, m^2_{E_i}) + C_{11}(...) + C_{12}(...)\right],\crn 
C^{\sigma^0_1}_{(32)R}&= \sum_{i=1}^3 \frac{-m_{\mu} Q_{E}Y^{\sigma^0_1*}_{1i}Y^{\sigma^0_1}_{2i}}{16\pi^2} \left[ C_2([p^2_i]; m^2_{\sigma^0_1}, m^2_{E_i}, m^2_{E_i}) + C_{12}(...) + C_{22}(...)\right],\crn 
C^{h_6}_{(32)L}&= \sum_{i=1}^3 \frac{-m_{\mu} Q_{E}Y^{h_6*}_{1i}Y^{h_6}_{2i}}{16\pi^2} \left[ C_2([p^2_i]; m^2_{\sigma^0_1}, m^2_{E_i}, m^2_{E_i}) + C_{12}(...) + C_{22}(...)\right],\crn 
C^{h_6}_{(32)R}&= \sum_{i=1}^3\frac{-m_{\tau}Q_{E}Y^{h_6*}_{1i}Y^{h_6}_{2i}}{16\pi^2} \left[ C_1([p^2_i]; m^2_{\sigma^0_1}, m^2_{E_i}, m^2_{E_i}) + C_{11}(...) + C_{12}(...)\right],\crn 
C^{h_6}_{(b1)L}&= \sum_{i=1}^3 \frac{-m_{e} Q_{E}Y^{h_6*}_{3i}Y^{h_6}_{(b-1)i}}{16\pi^2} \left[ C_2([p^2_i]; m^2_{\sigma^0_1}, m^2_{E_i}, m^2_{E_i}) + C_{12}(...) + C_{22}(...)\right],\crn 
C^{h_6}_{(b1)R}&= \sum_{i=1}^3\frac{-m_{b}Q_{E}Y^{h_6*}_{3i}Y^{h_6}_{(b-1)i}}{16\pi^2} \left[ C_1([p^2_i]; m^2_{\sigma^0_1}, m^2_{E_i}, m^2_{E_i}) + C_{11}(...) + C_{12}(...)\right],\crn 
C^{V}_{(32)L}&=-\frac{eg^2m_{\mu} Q_E}{16\pi^2 m^2_{V^0}}\sum_{i=1}^3V'^{E}_{1i}V'^{E*}_{2i} \crn 
&\times \left[ 2m_{V^0}^2 \left( C_0([p^2_i];m^2_{V^0}, m^2_{E_i}, m^2_{E_i}) + C_1(...) +2C_2(...) +C_{12}(...) +C_{22}(...)\right) \right. \crn 
&\left.  + m^2_{E} \left(-C_1(...) +C_{12}(...) +C_{22}(...) \right) +m_b^2 \left( C_1(...) +C_{11}(...) +C_{12}(...)\right) \right], \crn
C^{V}_{(32)R}&= -\frac{eg^2m_{\tau} Q_E}{16\pi^2 m^2_{V^0}}\sum_{i=1}^3V'^{E}_{1i} V'^{E*}_{2i} \crn 
&\times \left[ 2m_{V^0}^2 \left( C_0([p^2_i];m^2_{V^0}, m^2_{E_i}, m^2_{E_i})  +2 C_1(...) +C_2(...) +C_{11}(...) +C_{12}(...)\right) \right. \crn 
&\left.  + m^2_{E_i} \left(-C_2(...) +C_{11}(...) +C_{12}(...) \right) +m_a^2 \left( C_2(...) +C_{12}(...) +C_{22}(...)\right) \right], \crn 
C^{V}_{(b1)L}&=-\frac{eg^2m_e Q_E}{16\pi^2 m^2_{V^0}}\sum_{i=1}^3V'^{E}_{3i}V'^{E*}_{(b-1)i} \crn 
&\times \left[ 2m_{V^0}^2 \left( C_0([p^2_i];m^2_{V^0}, m^2_{E_i}, m^2_{E_i}) + C_1(...) +2C_2(...) +C_{12}(...) +C_{22}(...)\right) \right. \crn 
&\left.  + m^2_{E} \left(-C_1(...) +C_{12}(...) +C_{22}(...) \right) +m_b^2 \left( C_1(...) +C_{11}(...) +C_{12}(...)\right) \right], \crn
C^{V}_{(b1)R}&= -\frac{eg^2m_b Q_E}{16\pi^2 m^2_{V^0}}\sum_{i=1}^3V'^{E}_{3i}V'^{E*}_{(b-1)i} \crn 
&\times \left[ 2m_{V^0}^2 \left( C_0([p^2_i];m^2_{V^0}, m^2_{E_i}, m^2_{E_i})  +2 C_1(...) +C_2(...) +C_{11}(...) +C_{12}(...)\right) \right. \crn 
&\left.  + m^2_{E_i} \left(-C_2(...) +C_{11}(...) +C_{12}(...) \right) +m_a^2 \left( C_2(...) +C_{12}(...) +C_{22}(...)\right) \right], 
\end{align}
where $[p_i^2]=m_b^2,0,m_a^2$ relate to external momenta and the symbols $(...)$ stands for the list of arguments shown in the first terms. 
In the limit $m^2_a,m^2_b\simeq0$, the PV functions  $C_{0,i,ij}(0,0,0; m_B^2,m_F^2,m_F^2)$ are written as follows~\cite{Lavoura:2003xp}
\begin{align}
\label{eq_PVf}
C_{0}&= \frac{1-t +\ln(t)}{m_B^2(t-1)^2},\quad 
C_{1}=C_{2}=\frac{3 -4t + t^2+2\ln(t)}{4m_B^2(t-1)^3},\crn
C_{11}&=C_{22}=2C_{12} =\frac{11-18t +9t^2 -2t^3 + 6\ln(t)}{18m_B^2(t-1)^4}, 
\end{align}
where $t=m^2_F/m_B^2$.  Using these approximations we have 
\begin{align}
\label{eq_gsv}
g_s(t)&\equiv \left[ C_1 + C_{11}  + C_{12} \right] m_B^2  = \frac{t^3-6 t^2+3 t+6 t \ln (t)+2}{12  (t-1)^4},\crn 
g_v(t)&=2m_{V^0}^2 \left( C_0  +2 C_1 +C_2 +C_{11} +C_{12}\right)  + m^2_{E_i} \left(-C_2+C_{11} +C_{12}\right) \crn 
 & =\frac{-5 t^4+14 t^3-39 t^2+18 t^2 \ln (t)+38 t-8}{12 (t-1)^4}. 
\end{align}
These results are consistent with the formulas introduced in ref.~\cite{Freitas:2014pua}, used to discuss on the muon anomalous magnetic moments.  
\section{Equations for minimal conditions of the Higgs potential}
\label{app_minVhigg}
We have eight independent equations corresponding to eight neutral Higgs bosons $\left\{H_1^0, H_2^0,H_3^0,H_S^0,\sigma_1^0,\sigma_2^0, \sigma_S^0,\Delta_0\right\}$. 
In the limit of $\epsilon, k_2,k_S,n_1=0$  and the conditions in Eq.~\eqref{eq_laLimit} being applied, there are  seven independent equations that result in to the following functions:
\begin{align}
\label{eq_minEqHapp}
\mu_1^2&=  -2 k_1^2 \lambda _1^{\phi }+\frac{\sqrt{2} k_3 n_2 f^{\phi }}{k_1}-k_3^2 \lambda _{13}^{\phi }-n_2^2 \lambda _{12}^{\phi }-n_S^2 \lambda _2^{\phi S}, \crn 
\mu_{12}^2&=0,\crn 
\mu_3^2&= k_1^2 (-\lambda _{13}^{\phi })+\frac{\sqrt{2} k_1 n_2 f^{\phi }}{k_3}-2 k_3^2 \lambda _3^{\phi } , \crn 
f_{12}^{\phi S}&= 0 ,\crn
\mu^2_{22}&=  k_1^2 (-\lambda _{12}^{\phi })+\frac{\sqrt{2} k_1 k_3 f^{\phi }}{n_2}-2 n_2^2 \lambda _2^{\phi }-n_S f_{22}^{\phi S}, \crn 
\mu_S^2&=k_1^2 (-\lambda _2^{\phi S})-2 n_S^2 \lambda _1^S-2 n_S^2 \lambda _2^S  , \crn
f_{11}^{\phi S}&= 0.
\end{align}
Inserting them into the Higgs potential to cancel the dependent parameters, we can find the physical states and masses of the Higgs bosons as we discussed above. 

\end{document}